\documentclass{aa}
\usepackage[varg]{txfonts}
\usepackage[english]{babel}
\usepackage[utf8]{inputenc}
\usepackage[T1]{fontenc}
\usepackage{color, colortbl}
\usepackage{graphicx}
\usepackage{footnote}
\usepackage{txfonts}
\usepackage{natbib}
\usepackage{soul}
\usepackage{pifont}

\def\farcs{\hbox{$.\!\!^{\prime\prime}$}}

\newcommand{\appropto}{\mathrel{\vcenter{
  \offinterlineskip\halign{\hfil$##$\cr
\propto\cr\noalign{\kern2pt}\sim\cr\noalign{\kern-2pt}}}}}

\begin{document} 

\title{Oort Cloud comets discovered far from the Sun}

\titlerunning{Oort Cloud comets discovered far from the Sun}

\authorrunning{Królikowska and Dones} \author{Małgorzata Królikowska
  \inst{1}\fnmsep\thanks{\email{mkr@cbk.waw.pl}} \and Luke Dones
  \inst{2}\fnmsep\thanks{\email{luke@boulder.swri.edu}}
}

\institute{Centrum Badań Kosmicznych Polskiej Akademii Nauk (CBK PAN),
  Bartycka 18A, 00-716 Warszawa, Poland \and Department of Space
  Studies, Southwest Research Institute, 1050 Walnut St., Suite 300,
  Boulder, CO 80302, USA}

\date{Received 2023-06-14; accepted 2023-08-01}


\abstract
{Increasingly, Oort Cloud comets are being discovered at great
  distances from the Sun and tracked over ever wider ranges of
  heliocentric distances as observational equipment improves.}
{To investigate in detail how the original semimajor axis for
  near-parabolic comets depends on the selected data arc and the
  assumed form of the non-gravitational (NG) acceleration.}
{Among currently known Oort Cloud comets with large perihelion
  distances ($q > 3$~au), we selected 32 objects observed over the
  widest ranges of heliocentric distances in orbital legs before and
  after perihelion. For each of them, we determined a series of orbits
  using at least three basic types of data sets selected from
  available positional data (pre- and post-perihelion data and the
  entire data set), and a few forms of NG~acceleration representing
  water ice or CO sublimation.}
{We found that the motion of comets is often measurably affected by NG
  forces at heliocentric distances beyond 5\,au from the Sun. The most
  spectacular example is C/2010~U3 (Boattini), whose perihelion
  distance is 8.44~au.  NG effects are detectable for 19 of the 32
  comets within the positional data. For five comets, we found
  asymmetric effects of NG forces -- in three cases significantly
  greater before perihelion than afterward (C/2017~M4,
  C/2000~SV$_{75}$, and C/2015~O1), and in two others the opposite
  (C/1997~BA$_6$ and C/2006~S3).  We also find that the well-known
  systematic effect of finding more tightly bound original orbits when
  including the NG acceleration than in purely gravitational solutions
  may be related to the specific form of the standard $g(r)$ function
  describing the sublimation of ices.}
{}


\maketitle

\section{Introduction} 

Near-parabolic comets are increasingly discovered beyond Jupiter's
orbit and even past the orbits of Saturn and Uranus. We already have
some spectacular examples of these distant comets, such as C/2006~S3
(LONEOS), C/2010~U3 (Boattini), C/2014~UN$_{271}$
(Bernardinelli-Bernstein), and C/2017~K2 (Pan-STARRS). The Oort Cloud
comets discovered or seen in pre-discovery images before 2022 at
heliocentric distances of more than 10\,au are listed in
Table~\ref{tab:Oortspikecomets10au}. Four comets stand out on this
list because they are outside the orbit of Uranus ($\sim$~19~au) in
their pre-discovery images (column 4 in the table).

Of these comets, only two, C/2001~Q4 (NEAT) and C/2017~K2, have
perihelion distances $q<3.1$\,au. C/2017~K2 is unique in this group of
long-period comets (LPCs) due to its large heliocentric distance at
the time of first detection (see Fig.~5 in \cite{kroli-dyb:2019_stat}
and Fig.~1 in \cite{Meech_et_al:2017_17k2_co}); it passed perihelion
in December~2022. Recently, \citet{Yang_Jewitt:2021} detected carbon
monoxide (CO) gas from this comet at 6.7\,au from the
Sun. Supervolatiles such as CO and CO$_{2}$ have long been proposed to
drive the activity of some distant comets \citep{Meech2004,
  HarringtonPinto2022}, and CO-driven activity is possible in
principle even beyond 100~au \citep{Bouziani2022}.
\citet{Yang_Jewitt:2021} estimated that the outgassing from C/2017~K2
is about an order of magnitude smaller than the CO production rate
observed in C/1995~O1 (Hale–Bopp) at similar distances; see their
Fig.~3 and \citet{Womack2017, Womack2021}.

\begin{table*}
\caption{\label{tab:Oortspikecomets10au} A list of Oort Cloud comets
  discovered before 2022 more than 10~au from the Sun or with
  pre-discovery detections beyond 10~au. We list the comets'
  perihelion distances ($q$); their heliocentric distances at
  discovery and in their earliest pre-discovery detection; and their
  dates of perihelion passage $T_{\rm per}$. The perihelion date is
  given for the standard epoch close to the moment of perihelion
  passage.}  \setlength{\tabcolsep}{12.0pt} \centering
	\begin{tabular}{llrrr}
	\hline \hline 
Comet      & $q$  & \multicolumn{2}{c}{heliocentric distance [au]} & $T_{\rm per}$ \\

& [au] &  discovery   & pre-discovery                   & [yyyy\,mm\,dd]    \\
\hline
			C/2001 Q4  & 0.96 & 10.1  & --   & 2004\,05\,15 \\
			C/2003 A2  &11.4  & 11.5  & 11.9 & 2003\,11\,06 \\
			C/2005 L3  & 5.59 & 8.68\hspace{-0.2cm}  & 10.3 & 2008\,01\,16 \\
			C/2006 S3  & 5.13 & 13.3  & 26.1 & 2012\,04\,12 \\			
			C/2008 S3  & 8.01 &  9.46\hspace{-0.20cm} & 12.4 & 2011\,06\,10 \\
			C/2010 U3  & 8.48 & 18.4  & 25.8 & 2019\,02\,26 \\
			C/2012 LP$_{26}$  & 6.54  & 9.19\hspace{-0.2cm} & 10.0 & 2015\,08\,16 \\
			C/2014 B1  & 9.56 & 12.0  & --   & 2017\,09\,08 \\
			C/2014 UN$_{271}$ & \hspace{-0.2cm}10.9  & 29.0 & 34.1 & 2031\,01\,22 \\
			C/2015 D3  & 8.15 &  7.69\hspace{-0.2cm} & 10.5 & 2016\,05\,03 \\
			C/2016 Q2  & 7.08 & 12.6  & --   & 2021\,09\,28 \\
            C/2017 E3  & 5.92 &  5.95\hspace{-0.2cm} & 10.3 & 2017\,05\,31 \\
 			C/2017 K2  & 1.81 & 16.0  & 23.7 & 2022\,12\,19 \\
			C/2019 E3  &\hspace{-0.21cm}10.3  & 12.9  & 18.5 & 2023\,11\,15\\
			C/2019 U5  & 3.62 & 10.4  & 10.5 & 2023\,03\,29\\
			C/2020 F2  & 8.82 & 10.1  & 11.0 & 2022\,07\,15 \\
			C/2020 V2  & 2.23 & 8.71\hspace{-0.2cm}  & 10.1 & 2023\,05\,08 \\
			C/2021 A9  & 7.76 & 10.0  & 10.1 & 2023\,12\,01 \\
			C/2021 Q4  & 7.56 & 8.58\hspace{-0.2cm}  & 10.2 & 2023\,06\,10 \\
			C/2021 Q6  & 8.72 & 10.3  & 11.0 & 2024\,03\,21 \\
\hline
		\end{tabular}
\end{table*}

Among LPCs with smaller semimajor axes, a spectacular example of a
comet with a long data arc is C/1995~O1 Hale-Bopp, discovered in July
1995 7.1\,au from the Sun.  Its original semimajor axis was about
260\,au, while the comet is leaving the planetary zone on an orbit
with a semimajor axis of about 180\,au \citep{Marsden1997}. One
pre-discovery observation from April 1993, when the comet was 13.1\,au
from the Sun, was found. At the Minor Planet Center, more than 3,000
positional measurements for Hale-Bopp are listed. Hale-Bopp was
observed in August~2013 34.6~au from the Sun, outside Neptune's orbit,
and by JWST in July 2022 at 46.2~au \citep{Kelley2022}, making it the
only comet imaged well beyond Neptune's orbit. \citet{Szabo2012} find
that Hale-Bopp only stopped being active when it was about 28\,au from
the Sun, some 11 years past perihelion.

Starting in 2025, the Vera~C.~Rubin Observatory is scheduled to carry
out a ten-year program called the Legacy Survey of Space and Time
[LSST] \citep{Ivezic2019}. One of LSST's four science themes is to
take an inventory of small bodies in the solar system
\citep{Schwamb2023}, including long-period comets, in six filters,
with single-exposure limiting magnitudes of $r \sim 24.4$ and co-added
limiting magnitudes of $r \sim 27.5$. \citet{Silsbee2016} estimate
that LSST will discover hundreds to thousands of bodies from the Oort
Cloud with $q > 5$~au, even if they do not display cometary
activity. Given that some comets are active at $q \gg 5$~au, LSST may
find even more distant comets than \citet{Silsbee2016} predict.

\vspace{0.2cm}

One of the main goals of this study is to answer the question: How
does the value of the original semimajor axis $a_{\rm ori}$ depend on
the data-arc selection used for the orbit determination? Three main
types of data arcs were tested for each comet: complete arcs,
pre-perihelion arcs, and post-perihelion arcs.  This is an important
issue, as even tiny changes in the eccentricity of the near-parabolic
orbit can cause significant, systematic changes in the semimajor
axis. These systematic changes interfere with the estimates of the
position of the maximum of the original $1/a$~distribution even where
we might expect non-gravitational (NG) acceleration to be quite
negligible.

The dependence of $a_{\rm ori}$ on the data arc can only be studied
using the longest data arcs available. Therefore, we selected objects
observed for long intervals before and after perihelion passage,
enabling us to determine high-quality orbits based on a complete data
arc, as well as with observations taken only before or after
perihelion. We chose Oort Cloud comets with original semimajor axes of
more than 10,000~au according to a gravitational model of motion. We
also considered comet C/2016~N4 (MASTER) because its aphelion distance
is $\approx$~10,000~au (Sec.~\ref{sec:lpc_samples} and
\ref{sec:original-a}).

\vspace{0.2cm}

Another important objective of this study is to investigate the
non-gravitational (NG) effects in the orbits of LPCs with large
perihelion distances. \cite{squires:1961} and
\cite{marsden-sek-ye:1973} pointed out that there is a systematic
difference between gravitational (GR) and NG orbits for long-period
comets. \cite{marsden-sek-ye:1973} calculated GR and NG~orbits for
five LPCs with 0.3~au $\lesssim q \lesssim$~1.2~au, and found that in
general, NG~fits gave more tightly bound orbits. They also stated that
the explanation for this systematic effect ``is not
obvious". \cite{krolikowska:2001} calculated NG~orbits for sixteen
LPCs with 0.2~au $\lesssim q \lesssim$~2.4~au for which GR~orbit fits
gave hyperbolic original orbits (i.e., $1/a_{\rm ori} < 0$). In every
case, $1/a_{\rm ori}$ increased when NG effects were accounted
for. Fourteen of the orbits changed from hyperbolic to elliptical,
while the other two remained hyperbolic at the $\approx 2 \sigma$
level.

\cite{marsden-sek-ye:1973} introduced the ``$g(r)$'' functional form
(Eq.~\ref{eq:g_r}) for how the magnitude of NG~forces depends on
distance $r$ from the Sun (see Sec.~3). Most studies that have
calculated NG~orbits for comets, including orbital databases such as
the Minor Planet Center\footnote{\tt \tiny
https://www.minorplanetcenter.net/db\_search/} (MPC) and the
JPL~Small-Body Database Browser\footnote{\tt \tiny
https://ssd.jpl.nasa.gov/sbdb.cgi} (JPL), and \cite{krolikowska:2001}
use this equation, which is a fit to a simple energy balance model for
a spherical nucleus \citep{delsemme-miller:1971}. For very small $r$,
$g(r) \appropto r^{-2}$, while far from the Sun, $g(r)$ falls much
more rapidly with increasing distance. However, the distance $r_0$ at
which the transition between these two regimes takes place depends
upon which gas is the driver of activity and the rotation state of the
cometary nucleus (see Sec.~\ref{sec:orb_solution}).

The only comet for which we have {\em in situ} data for an extended
time (more than two years) is the target of the Rosetta mission, the
Jupiter-family comet 67P/Churyumov–Gerasimenko. For this comet, the
water production rate $Q(r) \appropto r^{-5}$ from $r \approx 4$~au to
perihelion at 1.24~au and $Q(r) \appropto r^{-7}$ after perihelion
\citep{Hansen2016, Farnocchia2021}.  In the model of
\citet{marsden-sek-ye:1973}, the magnitude of NG forces should be
nearly proportional to the gas production rate
\citep{Kramer2019}. Comet 67P has two main lobes, connected by a
``neck," and an obliquity of 52$^\circ$ \citep{Jorda2016}.  As a
result of its large obliquity, the comet spends most of its 6.4-year
orbit in northern summer. Southern summer lasts less than a year, but
perihelion passage occurs within this time. Most of the comet's mass
loss occurs during this brief period. \cite{Attree2019} invoke the
removal of a mantle from 67P's southern hemisphere to explain the
steep trend of $Q(r)$ vs.\ $r$. At least for this comet, modeling
seasonal effects is crucial to reconcile the different dependencies of
$g(r)$ and $Q(r)$ with $r$ \citep[also see][]{Davidsson2022,
  Attree2023}.

Unfortunately, we do not have a resolved image of the nucleus, much
less a shape model, for any long-period comet, and we only know the
spin states of a few LPCs \citep{Knight2023}. C/1996 B2 (Hyakutake),
an LPC with $a_{\rm ori} \approx 660$~au, has an obliquity of
108$^\circ$ and, like 67P, has asymmetric rates of gas and dust
production around perihelion due to seasonal effects
\citep{Schleicher2003}. Even for Hale-Bopp, the best-observed LPC, the
orientation of the spin axis is uncertain \citep{Jorda2002}.

From the attempts mentioned above to model comet activity in detail,
especially 67P, the need for a very individualized approach to
modeling comet activity emerges, as expressed by
\citet{Marschall2020}: ``Seasons and the nucleus shape are key factors
for the distribution and temporal evolution of activity and imply that
the heliocentric evolution of activity can be highly individual for
every comet, and generalizations can be misleading."

Modeling of the NG acceleration adjusted individually to both the
comet's astrometric data and observations of its activity is currently
possible only for a particular group of LPCs, such as the
near-parabolic comet C/2002~T7~(LINEAR)
\citep[$q=0.62$\,au,][]{kroli-dyb:2012}, where we have not only long
sequences of position data but also numerous measurements of the water
production rates \citep{Combi:2009}. Modeling the NG~motion using the
dedicated $g(r)$-form is especially possible in active comets which
are bright enough for the Solar Wind ANisotropies (SWAN) all-sky
hydrogen Lyman-alpha camera on the SOlar and Heliosphere Observer
(SOHO) satellite to detect and image at H~Ly$\alpha$ to calculate
water production rates. Many of these comets have perihelia well below
1\,au from the Sun, making them susceptible to partial or complete
disintegration, like the three comets studied by
\citet{Combi2023}. Since 1996 (when the bright comet
C/1996~B2~(Hyakutake) was intensively observed) to 2021, SWAN studied
about 50~LPCs \citep{Combi2019,Combi2021}; however, all have
perihelion distances below 2.0\,au, and about half are near-parabolic
(Oort Cloud) comets.

A long data arc covering a wide range of heliocentric distances
increases the chance of detecting NG~effects in large perihelion LPCs
active beyond 5~au, for which the sublimation of CO or CO$_{2}$ can be
suspected as the main driver of their activity \citep{Meech2004,
  HarringtonPinto2022}. In this aspect, it is partly a continuation of
research undertaken in \citet[][hereafter KD17]{kroli_dyb:2017}.  In
this paper, we will use the \citet{marsden-sek-ye:1973} formulation
for NG~effects, but will note issues with the model's parameters,
especially for CO-driven activity (Sec.~\ref{sec:orb_solution}).
 
This study also focuses on estimating how the original semimajor axes
of comets depend on the assumed form of comet activity with
heliocentric distance. We demonstrate what the differences can be by
applying a variety of simple models of activity for the studied comets
(Sec.~\ref{sec:2000sv_2015o1}--\ref{sec:other_examples}).
	
In the case of Oort Cloud comets with large perihelion distances
studied here, one would expect that the NG~effects would have a small
or negligible impact on the orbit determination, including the value
of the original $1/a$ obtained using the positional data.  In many
cases, we find a surprisingly large effect on the derived value of
$1/a_{\rm ori}$, as we discuss in Sec.~\ref{sec:original-a}.

Sec.~\ref{sec:conclusions} summarizes all aspects of this investigation.

\section{Samples of Long-Period Comets}\label{sec:lpc_samples}

\begin{table*}
	\caption{\label{tab:comet_list} A list of the analyzed sample
          of large-perihelion LPCs with long data arcs; three samples
          are listed in order of increasing $q$. We denote
          heliocentric distance as $r_h$. The final column gives
          $q_{\rm ori}$-number, the rank of the {\em original}
          perihelion distance, which we use in
          Figures~\ref{fig:LAn_gr}--\ref{fig:LBg_LBn}. This order is
          not identical to the order in osculating perihelion distance
          because $q$ changes for some comets during their passages
          through the planetary region. For instance, C/1980~E1 had an
          original perihelion distance of 3.17~au.  Samples LAn and
          LBn include comets with determinable NG~effects using
          astrometric data, where LAn includes comets with 3.1\,au\,$
          < q < 3.9$\,au and LBn those with $q > 4.5$\,au. The third
          sample, LBg, includes comets with $q > 4.5$\,au where
          NG~effects are not determinable with sufficient
          accuracy. `PRE' means pre-perihelion data arc, while `POST'
          means post-perihelion data arc. The six comets marked with
          asterisks in samples LAn and LBn have NG~effects determined
          for the first time by us. The two lines with different data
          arcs for C/2006~S3 and C/2009~F4 are described in the text.}
        \centering \setlength{\tabcolsep}{6.0pt}
	\begin{tabular}{llccccccrrc}
		\hline \hline
		Comet     & $q$    & first obs.   &  $T$  & last obs. & data & \multicolumn{2}{c}{range ~of ~r$_h$} & 
		\multicolumn{2}{c}{data arc}  \\
		&        & \multicolumn{3}{c}{}             & arc &  PRE     & POST      & PRE             & POST            & $q_{\rm ori}$- \\
		& [au]   & \multicolumn{3}{c}{[yyyymmdd]} & [yr] &  [au]   & [au]      & \multicolumn{2}{c}{[yr]/No. of obs.}& number \\
		\hline
		\multicolumn{10}{c}{\bf Sample LAn } \\   
		C/2000 CT$_{54}$ & 3.16 & 19990321 & 20010619 & 20040117 & 4.83 & 7.716~~--  &  8.519  & 2.25/69   &  2.58/152  & 1\\
		C/2016 N4        & 3.20 & 20150911 & 20170916 & 20200326 & 4.54 &  7.136~~-- &  6.376  & 2.01/1036 &  2.52/1049 & 3\\
		C/2017 M4*       & 3.25 & 20170616 & 20190118 & 20211130 & 4.46 &  6.088~~-- &  9.191  & 1.41/2854 &  2.83/430  & 4\\
		C/1980 E1        & 3.36 & 19800211 & 19820312 & 19861230 & 6.88 &  7.469~~-- & 13.91   & 2.02/97   &  4.78/101  & 2\\
		C/1997 BA$_6$    & 3.44 & 19970111 & 19991127 & 20040915 & 7.68 &  9.152~~-- & 13.39   & 2.84/278  &  4.54/258  & 5\\
		C/2012 F3        & 3.46 & 20120119 & 20150406 & 20171017 & 5.74 &  9.928~~-- &  8.359  & 3.20/1034 &  2.52/729  & 6\\
		C/1999 H3        & 3.50 & 19990422 & 19990818 & 20020320 & 2.91 &  3.664~~-- &  8.483  & 0.31/317  &  2.59/563  & 7\\
		C/2000 SV$_{74}$ & 3.54 & 20000905 & 20020430 & 20050512 & 4.68 & 6.264~~--  &  9.507  & 1.58/757  &  3.03/1450 & 8\\
		C/2015 O1        & 3.73 & 20150615 & 20180219 & 20200320 & 4.76 &  8.652~~-- &  7.276  & 2.68/1904 &  2.08/2260 & 9\\
		C/2013 G3        & 3.85 & 20130410 & 20141115 & 20161223 & 3.70 &  6.201~~-- &  7.337  & 1.58/850  &  2.08/62   & 10\\
		C/2005 EL$_{173}$& 3.89 & 20050303 & 20070305 & 20081117 & 3.71 & 7.120~~--  &  6.450  & 1.99/231  &  1.69/85   & 11\\
		\multicolumn{10}{c}{\bf Sample LBn} \\   
		C/2008 FK$_{75}$ & 4.51 & 20080331 & 20100929 & 20131106 & 5.60 &  8.239~~-- &  9.507  & 2.50/2388 & 3.10/1322 & 12\\
		C/1999 U4        & 4.92 & 19990918 & 20011028 & 20040411 & 4.56 &  7.552~~-- &  8.222  & 2.10/296  & 2.44/615  & 14\\
		C/2015 H2*       & 4.97 & 20140519 & 20160913 & 20191022 & 5.43 &  7.964~~-- &  9.519  & 2.31/91   & 2.97/115  & 15\\
		C/2006 S3        & 5.13 & 19991013 & 20120416 & 20201124 &21.1  & 26.1~~--   & 19.76   & 12.50/3183& 8.60/3386 & 16\\
		                 &      & 20060829 &          & 20180116 &11.3  & 14.40~~--  & 14.64   & 5.63/3180 & 5.75/3370 &   \\ 
		C/2013 V4*       & 5.19 & 20131023 & 20151007 & 20180806 & 4.79 &  7.352~~-- &  8.991  & 1.96/1370 & 2.82/1084 & 17\\
		C/2009 F4*       & 5.45 & 20090319 & 20111231 & 20150820 & 6.41 &  8.956~~-- & 10.55   & 2.78/618  & 3.63/595  & 19\\
		                 &      &          &          & 20150220 & 5.92 &            &  9.612  &           & 3.14/593  &   \\
		C/2005 L3*       & 5.59 & 20040716 & 20080116 & 20140109 & 9.48 & 10.30~~--  &  14.93  & 3.49/1559 & 5.96/3780 & 21\\
		C/2010 U3*       & 8.44 & 20051105 & 20190226 & 20210318 &15.36 & 25.75~~--  &  9.520  & 13.30/1882& 2.05/599  & 32\\
		\multicolumn{10}{c}{\bf Sample LBg} \\   
		C/2007 VO$_{53}$ & 4.84 & 20071020 & 20100426 & 20120720 & 4.75 &  8.314~~--  &  7.759  & 2.44/296  & 2.20/417  & 13\\
		C/2016 A1        & 5.33 & 20160101 & 20171123 & 20200215 & 4.12 &  7.305~~--  &  7.896  & 1.89/974  & 2.22/1195 & 18\\
		C/1999 N4        & 5.50 & 19980827 & 20000523 & 20020506 & 3.69 &  7.143~~--  &  7.490  & 1.73/170  & 1.94/175  & 20\\
		C/2010 R1        & 5.62 & 20100904 & 20120518 & 20140330 & 3.57 &  7.154~~--  &  7.401  & 1.70/793  & 1.87/826  & 22\\
		C/2002 J5        & 5.73 & 20010806 & 20030919 & 20060305 & 4.58 &  7.869~~--  &  8.439  & 2.12/334  & 2.46/273  & 23\\
		C/1999 F1        & 5.79 & 19990313 & 20020213 & 20050828 & 6.46 &  9.281~~--  &  10.38  & 1.85/112  & 2.92/56   & 24\\
		C/2010 S1        & 5.90 & 20100921 & 20130520 & 20150719 & 4.82 &  8.850~~--  &  8.025  & 2.66/5260 & 2.16/3344 & 25\\
		C/2017 E3        & 5.92 & 20131214 & 20170631 & 20200620 & 6.52 & 10.28~~--   &  9.53   & 3.45/152  & 3.04/480  & 26\\
		C/2012 K8        & 6.46 & 20120530 & 20140819 & 20160608 & 4.02 &  8.396~~--  &  7.811  & 2.22/668  & 1.80/319  & 27\\
		C/2012 LP$_{26}$ & 6.54 & 20120523 & 20150816 & 20181005 & 6.37 & 10.02~~--   &  9.865  & 3.23/196  & 3.14/246  & 28\\
		C/2015 XY1       & 7.93 & 20151204 & 20180502 & 20201020 & 4.88 &  9.525~~--  &  9.617  & 2.35/415  & 2.02/323  & 29\\
		C/2008 S3        & 8.01 & 20061227 & 20110603 & 20150717 & 8.55 & 12.36~~--   & 11.87   & 4.16/928  & 4.06/1025 & 30\\
		C/2015 D3        & 8.15 & 20130408 & 20160430 & 20200221 & 6.87 & 10.46~~--   & 11.92   & 3.06/163  & 3.78/151  & 31\\
		\hline
	\end{tabular}
\end{table*}

There are now quite a few comets observed for several years from a
heliocentric distance of at least 7\,au prior to perihelion passage to
a similar distance after perihelion.  From this group, we selected
32~LPCs whose perihelion distances exceed 3.1\,au and whose data arcs
are within the broadest heliocentric distance ranges possible
(relative to perihelion distance). Three were discovered closer to the
Sun than 7\,au: C/1999~H3 (LINEAR), C/2000~SV$_{74}$ (LINEAR), and
C/2013~G3 (Pan-STARRS); see Table~\ref{tab:comet_list}.

C/1980~E1~(Bowell) is the only comet in Table~\ref{tab:comet_list}
discovered before 1990. It suffered such strong planetary
perturbations (mainly due to an approach to within 0.228~au of
Jupiter) that it is leaving our solar system on a fast hyperbolic
trajectory ($v_\infty \approx 3.8$~km/s). At present, only
1I/'Oumuamua and 2I/Borisov have faster trajectories. The other five
objects were discovered in the last decade of the 20$^{\rm
  th}$~century, the rest only in the 21$^{\rm st}$~century. This
statistic reflects the increased ability to detect comets farther and
farther from the Sun. Table~\ref{tab:comet_list} shows these objects,
which we divided into three groups as follows.

\begin{itemize}
  
	\item Sample LAn consists of 11 comets with 3.1\,au\,$ < q <
          3.9$\,au that exhibit NG~effects in the orbital fits to the
          positional data. Most of these objects were described in
          KD17 and \citet[][hereafter KD20, also see the CODE
            Catalog\footnote{\tt \tiny
            https://pad2.astro.amu.edu.pl/comets/}]{kroli-dyb:2020_CODE}. For
          C/2017~M4 (ATLAS), the NG~orbits were determined in this
          study.
	\item Sample LBn includes 8 comets with $q>4.5\,$\,au that
          also demonstrate NG~effects in the orbital motion; five
          comets from this sample have NG~orbits obtained in this
          investigation (C/2005~L3, C/2009~F4, C/2010~U3, C/2013~V4,
          C/2015~H2).
	\item Sample LBg includes 13 comets with $q>4.5\,$\,au for
          which the NG~effects cannot be determined or are highly
          uncertain.
\end{itemize}

Table~\ref{tab:comet_list} also shows that for almost all the comets
in our samples, data start at heliocentric distances greater than
7\,au; for comets C/2006~S3 (LONEOS) and C/2010~U3 (Boattini),
pre-discovery observations were made well beyond the orbit of Uranus,
and for C/2014~UN$_{271}$, beyond the orbit of Neptune.  Likewise, all
but two of these comets have been tracked after perihelion to
heliocentric distances beyond 7\,au. In some comets with moderate
perihelion distances (sample LAn), the heliocentric range of the data
arc is narrower; however, they also have a relatively long data arc on
both legs of their orbits. The only exception is comet C/1999~H3
(LINEAR), which has a short data arc before perihelion (about 4
months). This comet and two others with less than 100 pre-perihelion
measurements (see C/1980~E1 and C/2000~CT$_{54}$ (LINEAR) in
Table~\ref{tab:comet_list}) will serve as examples of objects with
pre-perihelion-based orbits of poorer quality than the orbits of the
other comets analyzed here. (See the uncertainties of $1/a_{\rm
  ori,pre,GR}$ represented by the light blue solutions in
Fig.~\ref{fig:LAn_gr}.) Despite this, the full-data-based GR~orbits
are of the highest quality for all these comets.  Almost all are Oort
spike comets according to their GR~solutions ($a_{\rm
  ori,full,GR}>10,000$\,au); the exception is C/2016~N4, which has an
original semimajor axis of about 6,000\,au.

Throughout this paper, $1/a_{\rm ori}$ means the original reciprocal
of the semimajor axis and is given at a distance of 250\,au from the
Sun; for more discussion, see Sec.~6 in KD17, which also describes the
method for calculating the uncertainties of $1/a$ and other orbital
elements.  The other indices given in $1/a_{\rm ori}$ specify the type
of data arc taken for orbit determination (e.g., `pre' denotes
pre-perihelion data) and the type of model of motion (GR, NG).

Twelve of the comets studied here were identified in pre-discovery
images, often by routine sky surveys. Such comets can be seen by
comparing the date encoded in the comet's name to the date of the
first observation (given in column [3] of
Table~\ref{tab:comet_list}). Such early positional detection can
potentially be used to verify NG~accelerations in the motion of comets
with large perihelion distances.

\section{Orbital Solutions}\label{sec:orb_solution}

We not only attempted to use a complete data arc, but also to
determine the GR and NG~orbits of all comets considered here using
data before perihelion (pre-perihelion data; in short: PRE) and after
perihelion (post-perihelion data, POST), separately. The model of
motion and the methods used for orbit determination are described in
KD17 (and references therein), while the expressions we assume for the
non-gravitational acceleration are described below.

Thanks to the long data arcs, we were able to accurately determine
GR~orbits for the PRE and POST data separately for all the comets we
studied.

The Oort spike comet sample analyzed in KD17 included 16 comets with
determinable NG orbits. We reconsider eight of these here because they
have long data arcs and separate orbits for both orbital legs were
previously not determined: C/1980~E1, C/1997~BA$_6$, C/1999~H3,
C/2000~CT$_{54}$, C/2000~SV$_{74}$, 2005~EL$_{173}$, C/2006~S3, and
C/2008~FK$_{75}$; for C/2006~S3, the data arc is about 4\,yr longer
than we had previously. In all, we study 19~comets with NG~effects,
which form two sub-samples: LAn (3.1\,au\,$<q<$\, 3.9\,au, 11~comets)
and LBn ($q>4.5$\,au, 8~comets); see Section~\ref{sec:lpc_samples} and
Table~\ref{tab:comet_list}. The sample of large-perihelion ($q >
3.1$~au) Oort spike comets with determinable NG~orbits thus now
includes 27~objects.

For all the comets with NG orbits, we noticed some decrease of the RMS
error of the fit and reduction of trends in Observed Minus Calculated
[O-C] time variations when the NG model of motion was used for orbit
determination. KD17 and \citet{krolikowska:2020_NG} have detailed
discussions of this issue.

\subsection{Purely Gravitational Orbits. Comparison of Original $1/a$ with Other Sources}\label{GR_comparison}

\begin{table*}
	\caption{\label{tab:GRcomparison} Comparison of original $1/a$
          between different orbital sources (Sample~LBg). See the
          discussion in the main text for the MPC values of $1/a$ for
          the last three comets in the table, which are marked with
          asterisks. }  \setlength{\tabcolsep}{4.0pt} \centering
	\begin{tabular}{lrlccc}
		\hline \hline 
		Comet      & \multicolumn{5}{c}{Original ~~$1/a$ ~~in ~~ ~~units ~~of ~~au$_{-6}$.} \\
		&  \multicolumn{2}{c}{used in this paper} & MPC &  JPL & Nakano \\
		& $1/a$ &  $1/a$-sigma &  $1/a$ &  $1/a$ &  $1/a$ \\
		\hline
		C/1999 F1 (Catalina)          & 37.05 & 0.59  & 38.1  & 36.51  & 27 \\
		C/1999 N4 (LINEAR)            & 70.73 & 1.43  & 67.81 & 68.55  & 56 \\
		C/2002 J5 (LINEAR)            & 58.39 & 0.58  & 59.76 & 59.16  & 50 \\
		C/2007 VO$_{53}$ (Spacewatch) & 91.80 & 0.35  & 90.99 & 90.53  & 81 \\			
		C/2008 S3  (Boattini)         & 21.15 & 0.34  & 20.99 & 19.88  & 10 \\
		C/2010 R1 (LINEAR)            & 43.75 & 0.54  & 44.3  & 43.79  & 35 \\
		C/2010 S1 (LINEAR)            & 23.22 & 0.15  & 23.12 & 22.15  & 13 \\
		C/2012 K8 (Lemmon)            & 38.81 & 0.48  & 38.78 & 38.83  & 29 \\
		C/2012 LP$_{26}$ (Palomar)    & 40.06 & 0.49  & 41.96 & 42.04  & 31 \\
		C/2015 D3 (PANSTARRS)         & 27.91 & 0.80  & 27.63 & 27.22  & 15 \\
		C/2015 XY$_1$ (Lemmon)        & 30.07 & 0.58  & 34$^*$& 31.74  & 22 \\
		C/2016 A1 (PANSTARRS)         & 39.11 & 0.24  & 39$^*$& 38.33  & 28 \\
		C/2017 E3 (PANSTARRS)         & 36.37 & 0.36  & 39$^*$& 38.59  & 28 \\
		\hline
	\end{tabular}
\end{table*}

To highlight what differences can be expected for $1/a_{\rm ori}$ due
to the use of different methods for orbit determination from the
positional data, including the data treatment (their selection and
weighting), we compare the values we determine or were previously
reported in the CODE Catalog with three other well-known orbital
sources: the MPC, JPL and Nakano
Notes\footnote{http://www.oaa.gr.jp/\~{ }oaacs/nk.htm}. To make this
comparison, we choose comets from Sample LBg, for which we have GR
orbits from all these sources. We try to have orbits determined on the
basis of data arcs as similar as possible. That is, we take the most
recent orbit given in each source except for three comets taken from
the MPC (marked by asterisks in Table~\ref{tab:GRcomparison}). For
those comets, we use the second orbit given in the MPC database
because its value of $1/a_{\rm ori}$ is consistent with other MPC
solutions for those comets, but completely different from the
$1/a_{\rm ori}$ value of the first solution. We do not know the cause
of this discrepancy.  The data were retrieved from these three orbital
databases between 14 and 19 July 2023.

The MPC and Nakano Notes list values of $1/a_{\rm ori}$ together with
osculating orbital elements.  To obtain the $1/a_{\rm ori}$ of orbits
given by JPL, we used the JPL Horizons Web
Application\footnote{https://ssd.jpl.nasa.gov/horizons/app.html\#/}.

Table~\ref{tab:GRcomparison} shows that our values of $1/a_{\rm ori}$
are very close to JPL's values.  These differences never exceed
2.5\,au$_{-6}$\footnote{In this paper, we use the notation au$_{-6}$
to indicate $10^{-6}$~au$^{-1}$. For example, a comet with $1/a_{\rm
  ori}$ = 100~au$_{-6}$ has $a_{\rm ori} = 10^4$~au.}, and are usually
smaller. The same is true for comparisons between our values and MPC,
or MPC and JPL for almost all comets except one of three mentioned
above (C/2015 XY$_1$).  These differences are mainly due to the data
treatment -- the different criteria used for data selection and
whether or not weighting is applied to the data. For example, using a
slightly different data treatment, we obtain $1/a_{\rm ori}=66.83 \pm
1.62$~au$_{-6}$ for C/1999~N4 and $40.98 \pm 0.50$~au$_{-6}$ for
C/2012~LP$_{26}$. How data arc and the data treatment can change the
$1/a_{\rm ori}$ estimates can be also seen by looking at the MPC
database, where in most cases a series of orbital solutions are given
for individual comet. The interesting example represent there the
comet C/2017~E3 where six of eight values of $1/a_{\rm ori}$ given for
orbits obtained using different data arcs are in the range of
38-42\,au$_{-6}$. Using pre-perihelion data arc, we obtained a value
of $42.92\pm 1.04$\,au$_{-6}$.

We have noticed that the values of $1/a_{\rm ori}$ reported in Nakano
Notes are systematically lower by about 10~au$_{-6}$ than the values
from other sources. \citet{Ito2020} found a similar result, and
suggest that the discrepancy might result from assuming different
planetary masses; see \cite{Ito2021} for comparisons of $1/a_{ori}$
values in the CODE Catalog, JPL, and the MPC.

Below, we show that the differences in $1/a_{\rm ori}$ can be
10~au$_{-6}$ or even larger when we use only pre-perihelion part for
this sample of comets.

In this paper, when comparing the $1/a$ values for the studied comets,
we will mainly refer to those tabulated by the MPC, because at JPL,
these values are not given explicitly, but must be calculated by
setting the date for each comet to three or four centuries ago.

\subsection{Non-Gravitational Orbits}\label{subsec:NG_model}

To determine a cometary NG orbit in the region well inside the
planetary zone, as in KD17, we applied a standard formalism proposed
by \citet[][MSY]{marsden-sek-ye:1973}, where three orbital components
of the NG~acceleration acting on a comet are proportional to the
$g(r)$-function, which is symmetric relative to perihelion,

\begin{eqnarray}
F_{i}=A_{\rm i} \> g(r),& A_{\rm i}={\rm ~constant~~for}\quad{\rm i}=1,2,3,\nonumber\\
& \quad g(r)=\alpha(r/r_{0})^{-m}[1+(r/r_{0})^{n}]^{-k},\label{eq:g_r}
\end{eqnarray}

\noindent where $F_{1},\, F_{2},\, F_{3}$ represent radial,
transverse, and normal components of the NG~acceleration,
respectively, and the radial acceleration is defined as positive
outward along the Sun-comet line.  This study uses three types of the
$g(r)$-function defined above; their parameters are given in
Table~\ref{tab:gr-like_functions}.

\begin{figure}
	\centering
	\includegraphics[width=1.00\columnwidth]{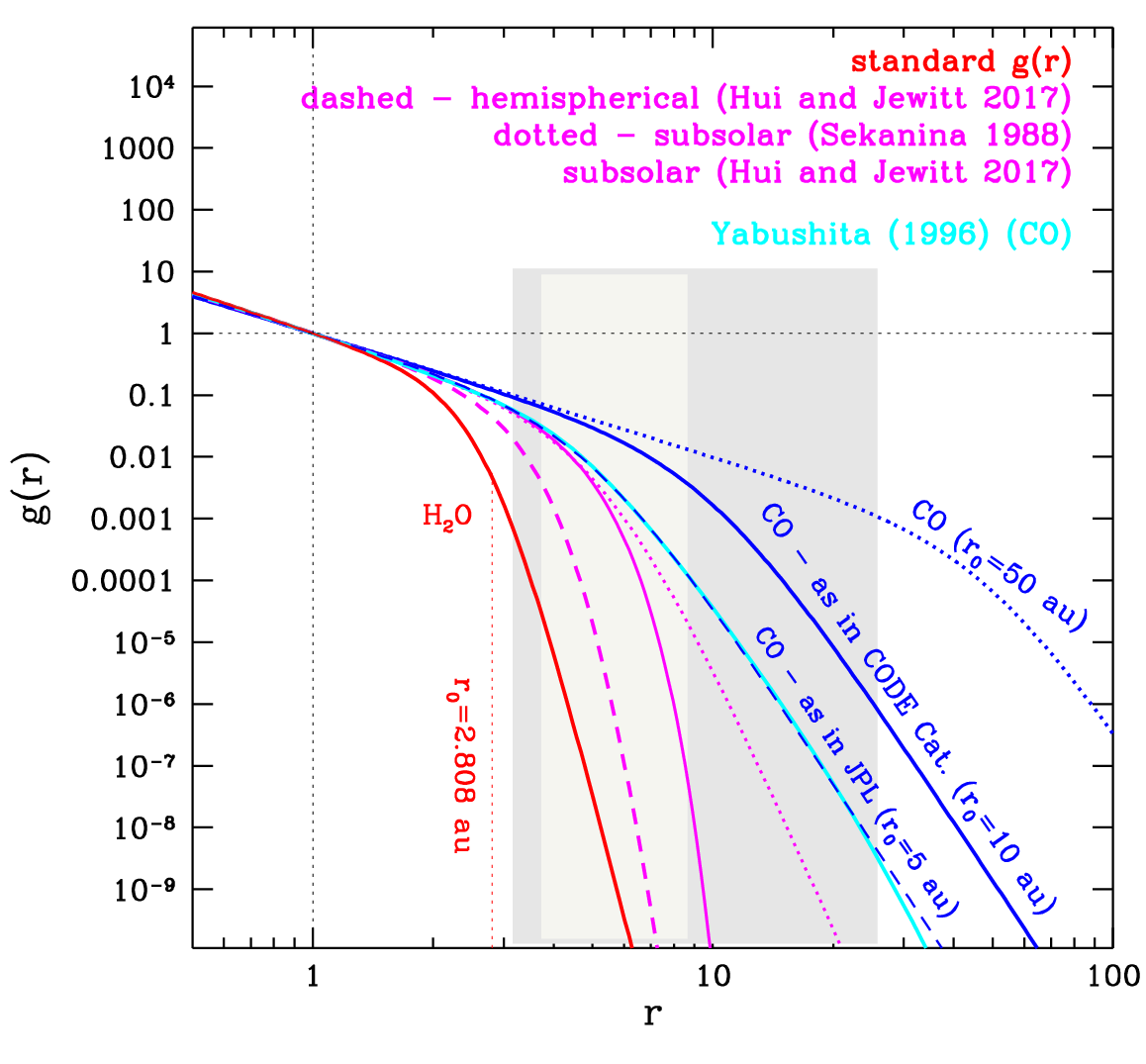} 
	\caption{Comparison of the $g(r)$-like formulas used in this
          study and given in the literature; logarithmic scales are
          used on both axes and $r$ is in au. The gray rectangle shows
          the range of heliocentric distances covered by all
          positional measurements used in this study (32 sets of
          data), while the light gray rectangle shows the range for
          C/2015~O1, chosen as a typical example for Sample LAn (see
          Table~\ref{tab:comet_list}). }
	\label{fig:grlike}
\end{figure}

The coefficient $\alpha$ is determined by the condition $g(1\,{\rm
  au}) = 1$. The values of $r_0$ for isothermal water-ice sublimation
and CO sublimation have a simple interpretation, as they represent the
approximate distance at which gas production rates are only $10^{-3}$
of the rates at 1~au. (To be precise, $g(r) = 10^{-3}$ at $r
\approx$~3.1~au and 10.9~au, respectively.) The value of $r_0 \approx
50$~au for subsolar water sublimation from HJ17 does not have the same
significance. In this case, $g(r) = 10^{-3}$ at $\approx 5.7$~au. One
should not attach special importance to the large values of $r_0$ and
$k$ given in HJ17 and used here. A similar very sharp cutoff for
$g(r)$ in Fig.~\ref{fig:grlike} can be obtained for many values of
$r_0$ and $k$.

Table~\ref{tab:gr-like_functions} shows that these $g(r)$-like
functions differ in the values of the exponents $m$, $n$, and $k$, but
most importantly, on the heliocentric distance range of the effective
action of the NG~acceleration, which is mainly controlled by the $r_0$
parameter \citep{Sekanina2021}. The standard form of the $g(r)$
function given by MSY, which we will call NG-std
(Table~\ref{tab:gr-like_functions}), is widely used to calculate the
NG~accelerations in comets and is shown in Fig.~\ref{fig:grlike} using
the red curve. Therefore, we applied this formula here to compare our
results directly with calculations in the literature. For models
involving water ice sublimation, we use the canonical value of
$r_0=2.808$\,au from \citet{marsden-sek-ye:1973} for a (roughly)
isothermal nucleus \citep{hui:2017_grlike}, and for CO ice
sublimation, we assume $r_0=10$\,au (as the basic CO form), and
$r_0=50$\,au (for comparisons) (and the same exponents $m = 2$, $n =
3$, and $k = 2.6$ used by JPL instead of the \citet{yabushita:1996_CO}
function used in \citet{krolikowska:2004}). The
\citet{yabushita:1996_CO} formula takes the form:
\begin{eqnarray}
  f(r)= 1.0006 \cdot r^{-2}\cdot 10^{-0.22185(r-1)/3} \cdot (1+0.0006r^{5})^{-1} \label{eq:Yab}
\end{eqnarray}
and is shown in Fig.~\ref{fig:grlike} using the cyan curve.  The
$g(r)$-like function with $r_0=10$\,au we use (blue curve in the
figure) is very similar in shape to Yabushita's function; however, our
curve has a knee at a larger heliocentric distance than does
Yabushita's formula or the $g(r)$-like formula used by JPL ($r_0 =
5$\,au).  More than 30 years ago, \citet{sekanina:1992} argued that
the knee is far outside the planetary zone when assuming sublimation
by CO from the subsolar point ($r_0 > 100$~au, his Fig.~3; also see
\citet{Meech2004}, \citet{Ye2020}, footnote 13, and
\citet{Bouziani2022}. Thus, the formula used here with the knee at
10\,au is a conservative estimate of the outer limit of cometary
activity. Furthermore, strong observational evidence has recently
emerged that some comets are active at the outskirts of the planetary
zone
\citep{Hui:2019a,jewitt:2021_distant_activity_2017k2,Bouziani2022}.
We will show that assuming $r_0 = 10$\,au (for CO sublimation) can
give significantly different results for original semimajor axes than
the standard $g(r)$ formula for water sublimation.
 
The standard $g(r)$ formula should only be treated as an approximation
of the actual momentum-transfer law in comets due to water-ice
sublimation. Recently, \citet[][HJ17]{hui:2017_grlike} discussed a
general approach to sublimation of a water ice comet nucleus and
determined three sets of parameters with isothermal, hemispherical,
and subsolar sublimation models, respectively (also see
\citet{Cowan1979})\footnote{These models differ only in the value of
the effective projection factor $\cos \zeta$ assumed for solar
illumination of the comet's nucleus (see Eq.~4 of HJ17). HJ17 set
$\cos \zeta$ equal to $\frac{1}{4}$, $\frac{1}{2}$, and 1 for,
respectively, the isothermal, hemispherical, and subsolar
cases.}. HJ17's $g(r)$-like function for isothermal sublimation is
very similar to that given by MSY (the lines almost overlap in the
plot, so we only show the curve representing $g(r)$-std formula). We
decided to use the MSY formula for the isothermal nucleus and HJ17 for
the subsolar sublimation of water ice (red and magenta solid lines in
the figure). The latter is similar for distances up to $\sim$5\,au to
the form found by \citet{sekanina:1988} for subsolar sublimation of
water ice; see Fig.~\ref{fig:grlike}. The difference between these two
curves (magenta and dotted magenta) at large heliocentric distances is
due to the different model assumptions of HJ17 and
\citet{sekanina:1988}.  For the formula representing the sublimation
of CO, we used the third and fourth sets of NG~parameters given in
Part~A of Table~\ref{tab:gr-like_functions}. As we already mentioned,
this function differs from the CO formula used in the JPL~Database
Browser by assuming $r_0 = 10$~au or larger instead of 5~au.  Such a
selection of $g(r)$-like functions to study the NG~orbits is necessary
because, to date, experience with modeling different sets of
parameters for $g(r)$ \citep{krolikowska:2004,kroli-dyb:2012}
indicates that the best sensitivity to the quality of NG~orbit fitting
to positional data can be found when testing different values of
$r_0$. Fig.~\ref{fig:grlike} shows that $r_0$ defines the heliocentric
distance of a knee in the $g(r)$-like curve \citep[see
  Figure~\ref{fig:grlike};][and KD17]{krolikowska:2004}. In the cases
of the magenta and blue curves with $r_0 = 5$ and 10\,au, these knees
are within the observation area (light gray band in the
figure). However, for the standard $g(r)$ function, this knee is less
than 3\,au from the Sun, therefore outside the gray area. Also, for
$r_0=50$\,au, the knee is outside the data area, beyond its right
edge. The different knee positions also result in substantially
various slopes of $g(r)$-like curves within the region covered by
observations.

\begin{table}
	\caption{\label{tab:gr-like_functions}Parameters used for the
          $g(r)$-like formula introduced in Eq.~\ref{eq:g_r}; for
          detailed discussion see text.}  \centering
        \setlength{\tabcolsep}{8.0pt}
		\begin{tabular}{ccccc}
\hline 			\hline 
			$\alpha$    & $r_0$ & $m$      & $n$   & $k$           \\
\hline \\
			\multicolumn{5}{c}{A.~~ $g(r)$ forms used in this study} \\ \\
			\multicolumn{5}{c}{Standard $g(r)$ function} \\ 
            \multicolumn{5}{c}{Isothermal water sublimation (MSY)}\\
			0.1113      & 2.808 & 2.15  & 5.093 & 4.6142     \\
			\\
			\multicolumn{5}{c}{Subsolar water sublimation (HJ17) }\\
			3.321 $\cdot 10^{-4}$  & 50.48 & 2.05  & 3.067 & 2752       \\
			\\
			\multicolumn{5}{c}{More volatile ices than water ice (CO sublimation)} \\
			0.01003     & 10.0  & 2.0   & 3.0   & 2.6        \\
			4.000 $\cdot 10^{-4}$  & 50.0  & 2.0   & 3.0   & 2.6        \\ \\
			\multicolumn{5}{c}{B.~~Other $g(r)$-forms shown in Fig.~\ref{fig:grlike}} \\ \\
			\multicolumn{5}{c}{Subsolar water sublimation (Sekanina 1988) }\\
            0.02727   & 5.6    & 2.1   & 3.2    & 3.9       \\
			\multicolumn{5}{c}{CO sublimation -- as in JPL }\\
			0.04084     & 5.0  & 2.0   & 3.0   & 2.6        \\
			\hline
		\end{tabular}
\end{table}

\subsection{Some Remarks on Orbit Fitting Assuming CO or H$_2$O-Driven Activity} \label{subsec:NG_model_CO}

CO ice is more volatile than water ice and is widely believed to be an
important driver of the activity in comets at large heliocentric
distances \citep[particularly beyond Saturn's
  orbit;][]{Bock_Morv:2004_book, AHearn:2012, Fulle:2020a,
  Yang_Jewitt:2021}. However, for comets with moderate perihelion
distances (3.1\,au\,$< q <$\,4.5\,au, LAn sample), we try three forms
of $g(r)$-like functions (MSY, HJ17, and CO sublimation with
$r_0=10$\,au) to compare the quality of the fits to the positional
data (see below). For comets with $q > 4.5$~au (samples LBn and LBg),
we only applied the $g(r)$-like function which describes the
sublimation driven by CO to NG~orbit determination.

In general, the relative uncertainties of the NG~parameters obtained
from the CO-driven $g(r)$-like formula are much smaller than those
obtained from the standard formula for comets observed at the
heliocentric distances analyzed here. The same is true for the
$g(r)$-like function given by HJ17 for subsolar water sublimation. It
seems to be a natural consequence of the much greater $r_0$ used for
the CO-driven and subsolar $g(r)$-like formulas because the
NG~acceleration then operates efficiently at much greater distances
from the Sun (within a longer part of the data arc) than in the case
of the standard $g(r)$-like formula with $r_0=2.808$\,au (see
Fig.~\ref{fig:grlike}).

For comets with $q > 3$~au, the position of the knee of the standard
form of the $g(r)$ function is closer to the Sun than their perihelion
distances. Typical values of the parameters $A_1, A_2$, and $A_3$ for
these comets are in the range of 10$^{-6}$ -- 10$^{-4}$
au$\cdot$day$^{-2}$. For the other forms of $g(r)$ (for which the knee
is much further from the Sun), the inferred values of $A_1, A_2$, and
$A_3$ are much smaller for the same comets.  Their typical values
range from 10$^{-9}$ to 10$^{-7}$ au$\cdot$day$^{-2}$. However, in all
cases the NG~accelerations at perihelion ($A\cdot g(q)$, where $A =
\sqrt{ {A_1}^2 + {A_2}^2 +{A_3}^2}$) are within the range $10^{-3}$ --
$10^{-5}$ of the solar gravitational acceleration $F_{\sun} (r)$ (see
next sections).

Two of the analyzed comets, C/2006~S3 and C/2010~U3, were discovered
much more than 10\,au from the Sun. Therefore, for these two comets,
we also test intermediate values of $r_0$ between 10\,au and 50\,au
for CO~sublimation (Secs.~\ref{subsec:2010u3} and
\ref{subsec:2006s3}).

\vspace{0.2cm}

\section{Sample LAn}\label{sec:2000sv_2015o1}

\begin{table*}
	\caption{\label{tab:NG_solutions} RMS errors of purely
          gravitational (GR) and non-gravitational (NG)~solutions for
          the full data arcs, and the PRE and POST data arcs for the
          LAn and LBn~samples of LPCs, in which NG~orbits can be
          determined using the entire data arcs shown in
          columns~[3]--[5] in Table~\ref{tab:comet_list}. Comets are
          listed in order of increasing osculating perihelion
          distance.  Sample LAn: The four values in columns [3], [5],
          and [7] with RMS refer to the solutions GR/NG$_{\rm
            std}$/NG$_{\rm CO}$/NG$_{\rm subsolar}$, respectively. In
          other cases, a description is provided in the preceding
          columns. The lack of any RMS value indicates that the
          solution is non-physical ($A_1<0$) or uncertain
          (uncertainties of $A_1$, $A_2$, and $A_3$ greater than the
          values themselves). For C/2015~O1, solutions based on
          unweighted data are also shown for comparison on the second
          line for this comet. Sample LBn: The three values of RMS
          describe the solutions GR/NG$_{\rm CO}$/NG$_{\rm
            CO,50\,au}$, respectively. In the case of
          C/2008~FK$_{75}$, we compare the results of KD17 (second
          line) with those redetermined in this study (first line);
          for more discussion, see Sec.~\ref{subsec:2008fk}.  }
        \centering \setlength{\tabcolsep}{3.0pt}
	\begin{tabular}{lcccccc}
		\hline			\hline 
		Comet        & \multicolumn{2}{c}{F U L L } & \multicolumn{2}{c}{P R E} & \multicolumn{2}{c}{P O S T} \\
		& type of  & RMS    & type of  & RMS   & type of  & RMS  \\  
		& solution &        & solution &       & solution &      \\    
		\hline
		\multicolumn{7}{c}{\bf Sample LAn} \\   
		C/2000 CT$_{54}$  &  & 0.81/0.73/0.74/0.73 & GR               & 0.52      & GR                & 0.68        \\
		C/2016 N4         &  & 0.51/0.42/0.43/0.43 & GR/NG$_{\rm CO}$ & 0.35/0.33 & GR/NG$_{\rm std}$ & 0.44/0.41   \\
		C/2017 M4         &  & 0.88/0.48/0.48/0.47 & GR/NG$_{\rm CO}$/NG$_{\rm sub}$ & 0.40/0.37/0.37 & GR/NG$_{\rm CO}$/NG$_{\rm sub}$  & 0.33/0.31/0.30 \\ 
		C/1980 E1         &  & 1.17/1.06/1.06/1.06 & GR & 1.14      & GR  & 0.98      \\ 
		C/1997 BA$_6$     &  & 0.74/0.67/0.70/0.65 &  GR/NG$_{\rm CO}$/NG$_{\rm subsolar}$  & 0.53/0.51/0.51 &   & 0.71/0.70/0.68/0.69 \\
		C/2012 F3         &  & 0.55/0.34/0.34/0.35 & GR/NG$_{\rm CO}$ & 0.48/0.46 & GR  & 0.53     \\
		C/1999 H3         &  & 0.67/0.50/0.51/0.51 & GR & 0.49 & GR/NG$_{\rm std}$ & 0.51/0.50/0.50/0.50   \\
		C/2000 SV$_{74}$  &  & 1.06/0.67/0.66/0.66 &    & 0.58/0.56/0.53/0.54 &    & 0.70/0.68/0.65/0.66 \\
		C/2015 O1         &  & 0.95/0.51/0.51/0.48 &    & 0.36/0.34/0.33/0.32 &    & 0.54/0.53/0.53/0.53 \\
		                  &  & 1.07/0.67/0.65/0.64 &    & 0.54/0.46/0.44/0.44 &    & 0.72/0.71/0.72/0.72 \\   
		C/2013 G3         &  & 0.47/0.38/0.38/0.38 & GR & 0.37  & GR  & 0.43     \\
		C/2005 EL$_{173}$ &  & 0.45/0.42/0.41/0.41 & GR & 0.43  & GR  & 0.36     \\
		\multicolumn{7}{c}{\bf Sample LBn} \\   
		C/2008 FK$_{75}$  &                   & 0.57/0.44/0.44   &                    & 0.42/0.40/0.40 &        & 0.49/0.48/0.48     \\
		                  & GR/NG$_{\rm CO}$  & 0.54/0.48        &     --             &   --           &  --    &     --             \\
		C/1999 U4         &                   & 0.72/0.70/0.70   & GR                 & 0.64           & GR     & 0.71               \\
		C/2015 H2         &                   & 0.72/0.34/0.35   & GR                 & 0.29           &  GR    & 0.38               \\
		C/2006 S3         &                   & 0.66/0.53/0.53   & GR/NG$_{\rm CO}$   & 0.52/0.51      & GR/NG$_{\rm CO}$ & 0.53/0.46\\
		C/2013 V4         &                   & 0.48/0.45/0.45   & GR                 & 0.44           & GR     & 0.44               \\
		C/2009 F4         &                   & 0.54/0.48/0.49   &                    & 0.33/0.32/0.32 & GR     & 0.60               \\
		C/2005 L3         &                   & 0.50/0.49/0.49   & GR                 & 0.37           &        & 0.53/0.52/0.52     \\
		C/2010 U3         & GR/NG$_{\rm CO}$  & 0.59/0.58        & GR/NG$_{\rm CO}$   & 0.56/0.54      &  GR    & 0.55               \\
		\hline
	\end{tabular}
\end{table*}

As we mentioned in the previous section, for comets having perihelion
distances between 3.1\,au and 4.5\,au, we fit three basic forms of the
$g(r)$-like function described in Part~A of
Table~\ref{tab:gr-like_functions} to the complete data arc, as well as
to the PRE and POST legs of their orbits, independently. However, we
tested the formula with $r_0=50$\,au for only four comets from this
sample (C/2017~M4, C/2000~SV$_{74}$, C/2015~O1 and C/2005~EL$_{173}$,
see below).

Sample~LAn contains six comets discovered before 2010. Their NG orbits
based on complete data arcs were considered in KD17. For the five
remaining objects (discovered after 2010), the NG~orbits are given in
the CODE~Catalog; however, longer data arcs are now (as of July~2023)
available for three of them (C/2016~N4, C/2017~M4, C/2015~O1). The
NG~orbits obtained for all of these comets using the full data arcs
show a reduction in the RMS and a better time distribution of the
residuals of the [O-C] diagram compared to the GR orbits; see
column~[3] of Table~\ref{tab:NG_solutions}. The most spectacular RMS
decreases in the NG~solutions were observed for C/2017~M4,
C/2000~SV$_{74}$, and C/2015~O1; these are discussed in more detail
below. NG~orbits based on the CO-driven formula obtained by
\citet{yabushita:1996_CO} are available in the Nakano Notes for all
three comets, as well as for C/1997~BA$_6$. The expression used by
Nakano is Eq.~4.4 of \cite{yabushita:1996_CO} (in different units) and
is given in Eq.~\ref{eq:Yab} in the previous section and is shown in
Fig.~\ref{fig:grlike} with a cyan curve.  The NG~orbit of C/2015~O1 is
also given at the MPC.  JPL offers NG~orbits for C/2017~M4 and
C/2015~O1. Both sources and the Nakano Notes use different forms of a
$g(r)$-like function\footnote{JPL uses the standard $g(r)$ function
for C/2017~M4, whereas for C/2015~O1, they use the same values of $m$,
$n$, and $k$ for CO sublimation that we list in
Table~\ref{tab:gr-like_functions}, but set $r_0 = 5$~au, which implies
$\alpha = 0.040837$; see Part B of Table~\ref{tab:gr-like_functions}.
Fig~\ref{fig:grlike} shows that this $g(r)$-like formula for CO is
almost the same as Yabushita's formula.}.  It is also worth
highlighting the strong NG~effects in the positional data of comet
C/2012~F3, which were previously discussed in \citet[][also see the
  CODE Catalog]{wys-dyb-kro:2020_first_stars}; those orbits based on
the complete set of data are also used here. The weakest evidence of
NG~effects was found in the positional data of comet
C/2005~EL$_{173}$.

Table~\ref{tab:NG_solutions} also shows that it was possible to obtain
independent NG~solutions for the PRE or POST data arcs in half of the
cases, but only for four comets for all forms of $g(r)$ on both
orbital legs.

Table~\ref{tab:NG_solutions} shows that the RMS values for
NG~solutions (water ice sublimation and CO-driven) obtained using the
full data arcs (column [3]), as well as those based on PRE and POST
data arcs (columns [5] and [7]), are very similar.  Moreover, for some
comets with NG~solutions determinable for all three types of data
sets, the preferred form of $g(r)$ seems to be different depending on
the choice of the data arc set.  These facts make it difficult to draw
firm conclusions.

However, considering all three types of data arcs and also time trends
in residuals ([O-C]), we can say that for three comets
(C/2000~CT$_{54}$, C/2016~N4, and C/1999~H3), the orbit based on the
standard form of water ice sublimation gives a slightly better fit
than does the CO-driven formula. In the other five cases (C/2017~M4,
C/1997~BA$_6$, C/2000~SV$_{74}$, C/2015~O1, and C/2005~EL$_{173}$),
the NG~solutions based on the sublimation of CO ice or sublimation of
water ice from the subsolar point appear to fit the data better than
the NG~solution using the standard $g(r)$. For the next two comets
(C/2012~F3, C/2013~G3), all three types of NG~solutions have RMS
values within 0.01~arcsec of each other.

For C/1997~BA$_6$ and C/2015~O1, the NG~orbit obtained using the
subsolar water sublimation formula gives the smallest RMS of the three
NG~solutions considered here.  In addition, NG~solutions with
asymmetric $g(r)$-like formulas (i.e., with the magnitude of the NG
forces peaking before or after perihelion) can also be obtained in the
last two cases. Of these asymmetric NG~solutions, the orbit assuming
the sublimation of water ice from the subsolar point also gives the
best fit, with RMS values of 0\farcs 65 and 0\farcs 46, respectively,
for C/1997~BA$_6$ and C/2015~O1.

For comets with 3.1\,au~$\leq q \leq$~4.5\,au, activity can be driven
by H$_2$O, CO, or another volatile (Fig.~\ref{fig:grlike}). One might
expect that hypervolatile ices like CO would be depleted in returning
long-period comets, but \citet{HarringtonPinto2022} found that LPCs
with $a_{\rm ori} > 10,000$~au typically produce more CO$_2$ than CO,
while LPCs with $a_{\rm ori} < 10,000$~au produce more CO than CO$_2$
(also see \citet{Womack2017, Womack2021} for discussions of the
dynamically old LPC C/1995 O1
(Hale-Bopp)). \citet{HarringtonPinto2022} invoke preferential loss of
CO from Oort Cloud comets due to galactic cosmic rays as a possible
mechanism for this trend \citep{Strazzulla1991, Maggiolo2020,
  deBarros2022}.

Thus, it seems that a CO $g(r)$ formula should be applied if we have
direct observations of the domination of CO-driven activity in a
specific object. Otherwise, we should try both H$_2$O- and CO-based
formulas.  At the moment, we do not have an updated analysis of the
dynamical status of these comets. For example, estimates in the CODE
Catalog should be updated in the future because more precise
parameters for stars that passed near the Sun in the recent past have
become available thanks to the {\it Gaia} Early Data Release~3
\citep{GaiaEDR3-summary:2021} and Data Release~3
\citep{GaiaSummary2023}; for more discussion, see \citet{Dyb-Bre:2021}
and \citet{Bailer-Jones2022}).

However, NG~orbits based on the CO-driven formula used here often seem
to reduce the trends in [O-C] (compared with the GR~orbits) and yield
smaller uncertainties for the NG~parameters than do NG~orbits based on
the two water-driven formulas. Moreover, the standard $g(r)$ formula
sometimes gives an unphysical negative radial component of the NG
acceleration (i.e., $A_1 < 0$) for the PRE or POST orbital leg.
Therefore, we ultimately decided to discuss the PRE/POST asymmetry in
NG~activity for C/2017~M4, C/1997~BA$_6$, C/2000~SV$_{74}$, and
C/2015~O1 using NG~orbits based on the CO-driven $g(r)$-like form, but
in Sec.~\ref{subsec:2017m4}--\ref{subsec:2015o1}, we also show the
asymmetry of activity for all NG~solutions given in
Table~\ref{tab:NG_solutions}.

\subsection{C/2017~M4 (ATLAS)} \label{subsec:2017m4}

C/2017~M4 was observed for 4.5\,yr in the heliocentric distance range
of 6.09\,au -- 3.25\,au (perihelion) -- 9.19\,au.  We were able to
obtain GR and NG~orbits for this comet for all three data arcs
considered (see Table~\ref{tab:NG_solutions}); the NG~parameters are
presented in Table~\ref{tab:NG_parameters_LAn_97ba_15o1}.

NG~orbits are fit to the entire data arc, resulting in an almost
twofold smaller RMS and a notable reduction of trends in [O-C]
compared to the GR~solution. When the PRE or POST orbital branch is
separately considered, the RMS only decreases a little for the NG
solutions, as the RMS for GR orbits is already very small. We observed
a similarly small decrease in the RMS values between GR and NG
solutions for the PRE or POST data arcs in all the other comets we
considered (see Table~\ref{tab:NG_solutions}).

The NG~parameters obtained here based on the standard $g(r)$ and full
arc are consistent with those obtained by JPL for the same data arc
(as of July 2023). JPL's values are: $A_1 = 903.4\pm 21.9$, $A_2 =
108.3\pm 24.6$, and $A_3 = 215.8\pm 8.1$ in units of
$10^{-8}$\,au$\cdot$day$^{-2}$ (compare with
Table~\ref{tab:NG_parameters_LAn_97ba_15o1}). However, for this comet,
we found that the NG~solutions for PRE and POST data sets based on
$g(r)$-std are unphysical or uncertain. Moreover, we obtain a slightly
better fit to the data when we use the other two $g(r)$-like
formulas. For both of these NG~functions, we also obtained reasonable
NG~parameters for the PRE and POST parts of the orbit separately. The
level of activity measured by $A = \sqrt{ {A_1}^2 + {A_2}^2 +
  {A_3}^2}$ is almost twice as high for the PRE branch than for the
POST branch for this comet when using a $g(r)$-like function for CO,
and about 30\% higher for HJ17; see
Table~\ref{tab:NG_parameters_LAn_97ba_15o1}. Thus, the relative
difference in NG~activity depends on the assumed form of the
$g(r)$-like function.

For this comet, we obtained different values of $1/a_{\rm ori}$,
depending on the model of motion and choice of the data arc. Taking
into account all GR and NG~orbits\footnote{In our discussions of the
range of original $1/a$ for C/2017~M4 and other comets, we do not
consider solutions with $r_0=50$\,au.} obtained here and based on the
PRE and entire data arcs, we found values of $1/a_{\rm ori}$ between
$14$\,au$_{-6}$ and $56$\,au$_{-6}$; each of the $1/a_{\rm ori}$
values has a small uncertainty of at most 2\,au$_{-6}$. Such small
uncertainty for individual solutions is the result of the highest
quality of all these orbits.  The NG~solution given by the Nakano
Notes (NK\,4608) gives a value of $1/a_{\rm ori}$ = 31\,au$_{-6}$,
whereas the GR orbit presented by the Minor Planet Center (MPC) gives
39\,au$_{-6}$ (data arc only through March 20, 2020).

We also tested the CO formula for $g(r)$ with $r_0=50$\,au for this
comet. This formula gives the same RMS within 0\farcs 01 as the CO
formula with $r_0=10$\,au and small differences in [O-C] with no
explicit improvement for all three types of data arcs (entire data
set, PRE, and POST). However, for the PRE data set the transverse
NG~parameter, A$_2$, is uncertain due to its small value and the
original $1/a$ is negative (see Sec.~\ref{sec:original-a}).

\subsection{C/2000~SV$_{74}$ (LINEAR)}\label{subsec:2000sv74}

\begin{figure*}
	\centering
	\includegraphics[width=1.00\columnwidth]{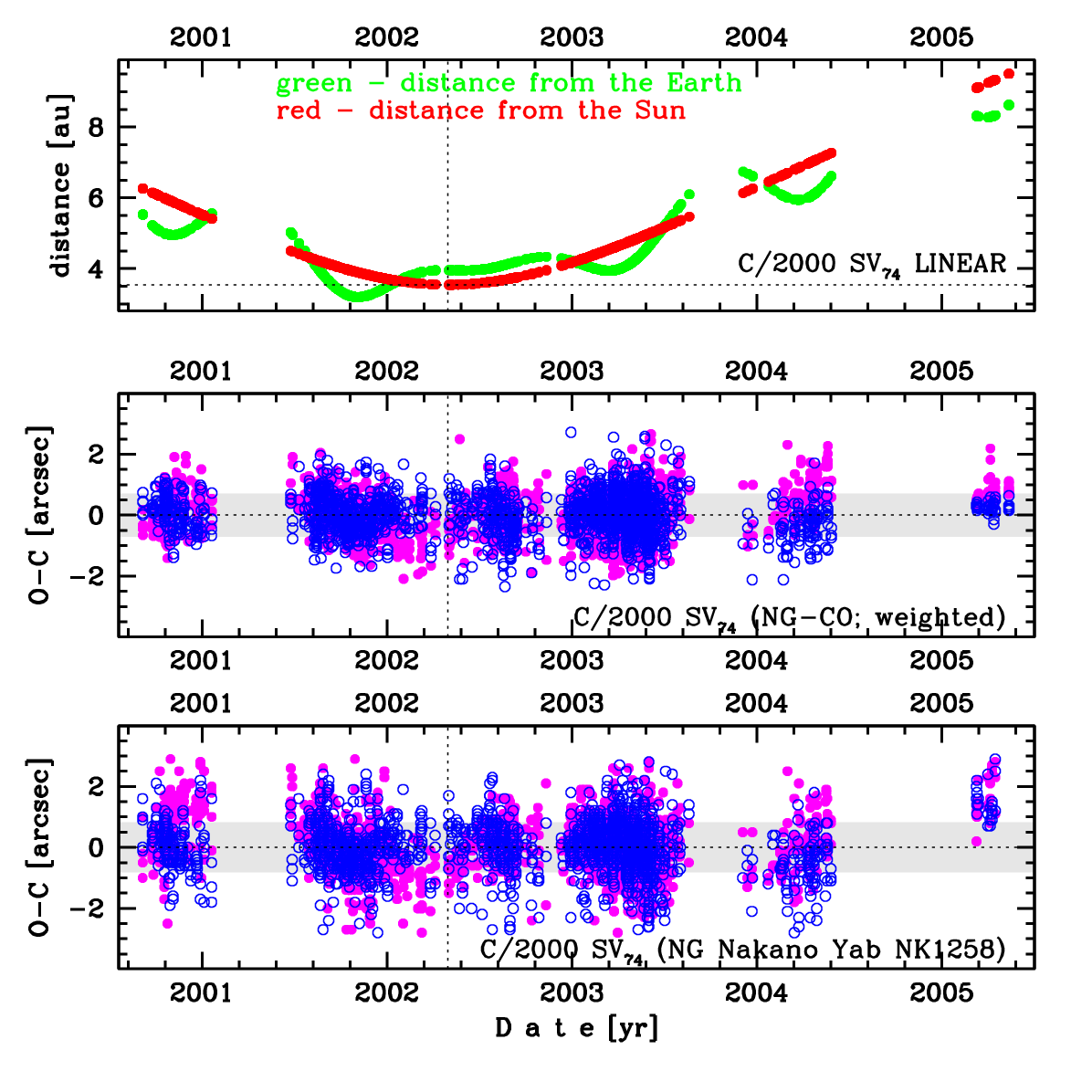}
	\includegraphics[width=1.00\columnwidth]{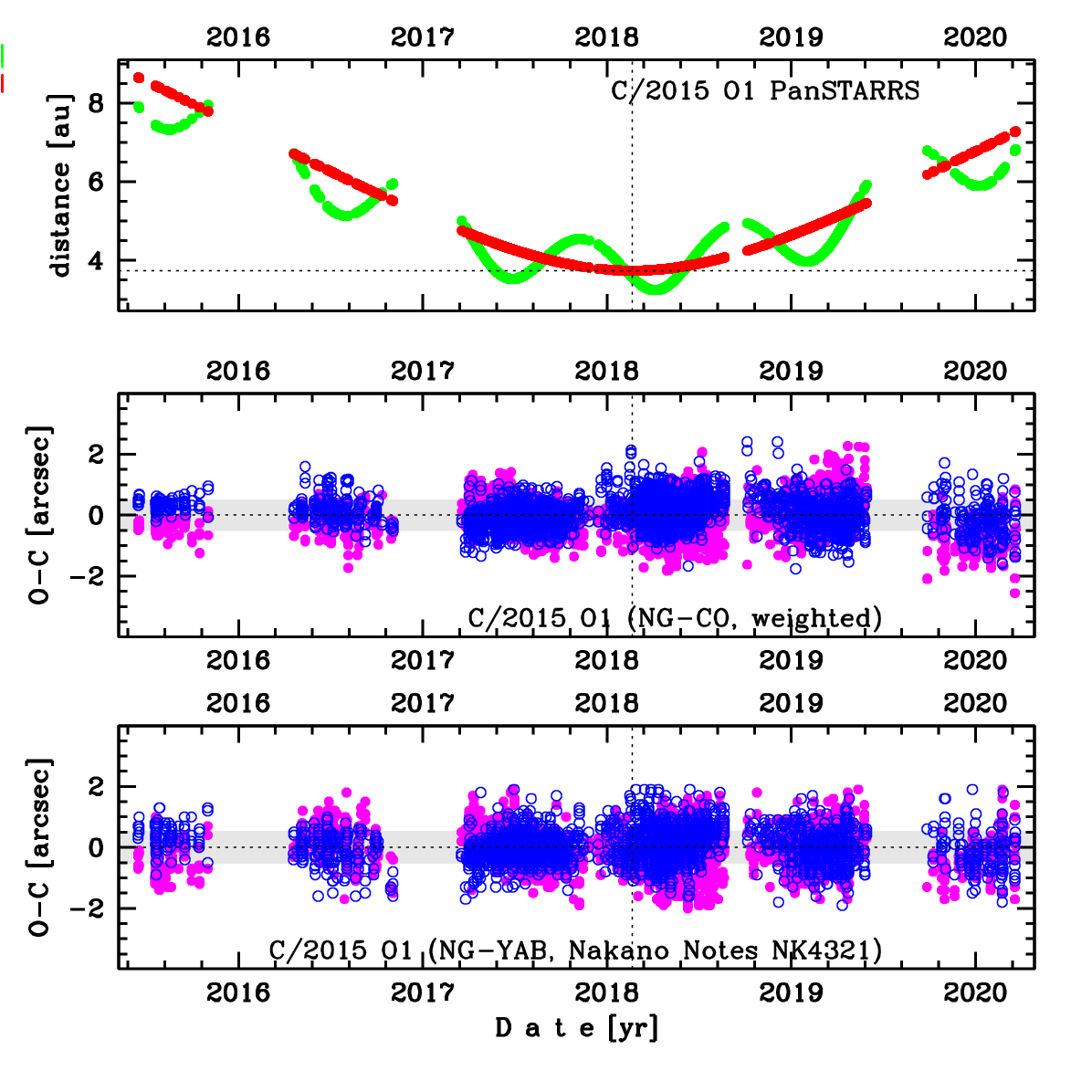}
	\caption{Comets C/2000~SV$_{74}$ (LINEAR) (left column) and
          C/2015~O1~Pan-STARRS (right column). Upper panels: Time
          distribution of positional observations with corresponding
          heliocentric (red curve) and geocentric (green curve)
          distances at which they were taken. The horizontal dotted
          line shows the perihelion distance for a given comet,
          whereas the vertical dotted line shows the moment of
          perihelion passage. Middle panels: [O-C]-diagrams for the
          NG~orbit based on the CO-driven $g(r)$-like function and
          obtained using the full data arc, where residuals in right
          ascension are shown using magenta dots and in declination by
          blue open circles; gray horizontal bands around zero
          indicate the RMS level. Bottom panels: [O-C]-diagrams using
          residuals available at the Nakano Notes Database and
          obtained using the \cite{yabushita:1996_CO} form of the
          CO-driven $g(r)$-like function.}
	\label{fig:c2000sv_nakano}
\end{figure*}

C/2000~SV$_{74}$ was observed for 4.7\,yr in the heliocentric distance
range of 6.26\,au -- 3.54\,au (perihelion) -- 9.51\,au; pre-discovery
observations extend the data arc by 19~days (see
Table~\ref{tab:comet_list}).  The CODE
Catalog\footnote{https://pad2.astro.amu.edu.pl/comets/orbit.php?\&int=2000sva5\&orb
=osculating} shows a significant decrease in RMS for the NG~orbit and
a notable reduction of trends in [O-C] compared to the GR~solution.

Fig.~\ref{fig:c2000sv_nakano} compares the NG~solution (based on the
CO-driven $g(r)$-like formula) obtained using the complete data arc
(with a slightly different weighting of observations than in the CODE
Catalog, hence there are tiny differences with the NG solution given
in the catalogue, see Table~\ref{tab:NG_parameters_LAn}) to the
NG~solution provided by the Nakano Notes (NK\,1258)
using the formula of \citet{yabushita:1996_CO}. The residuals taken
from the Nakano Notes are shown in the bottom panels of
Fig.~\ref{fig:c2000sv_nakano}. It seems that there are no fundamental
differences in the fits for both NG~solutions using different forms of
the expressions of the non-gravitational force for ices more volatile
than H$_{2}$O, except for the first and last oppositions for this
comet. Perhaps the CO formula we use in this study is a little better
in both these oppositions, but doubts remain about the quality of
these fits. A satisfactory pattern of residuals for the first and last
oppositions can be easily obtained by fitting two independent NG
orbits to the PRE and POST data and applying any of the three
$g(r)$-like formulas used here. Moreover, NG~solutions based on the
CO-driven and subsolar water sublimation formulas give RMS values for
both orbital legs a bit smaller than does the standard $g(r)$ formula,
see Table~\ref{tab:NG_solutions}.

Table~\ref{tab:NG_parameters_LAn} compares solutions based on HJ17 and
CO sublimation; however, the NG~solutions using HJ17 give slightly
worse RMS values for each of the three data arcs than does CO, and
thus is not listed in Table~\ref{tab:NG_solutions}.  They are shown
here to indicate similar relations between the NG~parameters on the
PRE and POST orbital legs.  For the NG~solution based on CO
sublimation, the level of activity before perihelion is almost twice
as large as after perihelion, when the asymmetry is measured by
comparing the values of $A$ (Sec.~\ref{subsec:2017m4}).  For the NG
solution based on the subsolar water sublimation model, activity is
about 2.5 times larger before perihelion than after perihelion and is
even larger for the standard form of $g(r)$; however, this last form
gives a worse fit to positional data for each set of data (full, PRE,
POST).

For this comet, we obtained different values of $1/a_{\rm ori}$,
depending on the model of motion and choice of which orbit is the most
reliable. Considering GR and NG~orbits obtained here (based on PRE and
complete data arcs), we conclude that the value of $1/a_{\rm ori}$ is
between $48$\,au$_{-6}$ and $91$\,au$_{-6}$; each of the $1/a_{\rm
  ori}$ values has a small uncertainty of at most 2\,au$_{-6}$.

Unfortunately, we obtained a similarly wide range of $1/a_{\rm ori}$
values for NG orbits because $1/a_{\rm ori}$ strongly depends on the
adopted $g(r)$-like formula for this comet. However, if we decide that
NG~orbits based on the CO-driven formula are the most reliable, we get
a range from 51\,au$_{-6}$ to 68\,au$_{-6}$, and this spread of
$1/a_{\rm ori}$ values is mainly due to the use of different data arcs
(full and PRE); the NG~solution in NK\,1258 gives a value of $1/a_{\rm
  ori} = 74\,$au$_{-6}$.

As for C/2017~M4, we also tested the CO formula with
$r_0=50$\,au. This $r_0$ gives a slightly smaller RMS of about 0\farcs
01, compared with the CO formula with $r_0=10$\,au for the POST data
set; however, no important differences in [O-C] were noticed for all
three types of data arcs (entire data set, PRE, and POST); see also
Sec.~\ref{sec:original-a}.

\subsection{C/2015~O1~(Pan-STARRS) and a Remark about C/1997 BA$_6$~(Spacewatch)} \label{subsec:2015o1}

	\begin{table*}
	  \caption{\label{tab:NG_parameters_LAn_97ba_15o1}
            NG~parameters $A_1$, $A_2$, and $A_3$ (in units of
            $10^{-8}$\,au$\cdot$day$^{-2}$) for chosen NG~orbits
            described in Table~\ref{tab:NG_solutions} for C/2017~M4
            (Sec.~\ref{subsec:2017m4}), C/1997~BA$_{6}$ (with a few
            more observations than in the CODE Catalog) and C/2015~O1
            (Sec.~\ref{subsec:2015o1}), all from the sample LAn. For
            all comets, the smallest RMS is for the NG~orbit based on
            HJ17 and using the entire data arc. For the PRE and POST
            solutions, NG~parameters for NG-CO and HJ17 are shown as
            in Table~\ref{tab:NG_solutions} (because all remaining
            NG~solutions have poor accuracy (marked as `un' in the
            column for $A_1$) or negative radial NG~parameter
            ($A_1$)). The last column gives the asymmetry in NG
            parameters between the PRE and POST orbital legs; for more
            discussion, see text. }
                 
	\setlength{\tabcolsep}{3.5pt} 
	\centering
	\begin{tabular}{lrrrrrrrrrrr}
		\hline 	\hline 
		Type of        & \multicolumn{3}{c}{f u l l ~~~a r c} & \multicolumn{3}{c}{P R E} & \multicolumn{3}{c}{P O S T} & References & Asymm.\\ 
		NG~model       & A$_1$   & A$_2$   & A$_3$   & A$_1$   & A$_2$   & A$_3$     &  A$_1$   & A$_2$   & A$_3$       &        & PRE/POST \\ 
		\hline 
		\\
			\multicolumn{12}{c}{ Comet C/2017~M4} \\    
NG-std   & $946.6$       & $57.20 $     & $217.7$      & $< 0$ &  &  & un &  &  & here  &  --  \\    
& $\pm 9.7$   & $\pm 11.91$ & $\pm 3.80$   &  &  &  &  &  &  &   & \\  %
\\
NG-subsolar     & $4.979$      & $0.3270 $    & $0.9932$      & 7.625        & $-9.569$     &  0.4456 & 7.606        & $-3.118$     & 1.169  & here  & 1.47\\
water       & $\pm 0.056$  & $\pm 0.0698$ & $\pm 0.0180$   & $\pm 0.499$  & $\pm 0.575$  & $\pm 0.2054$& $\pm 0.664$  & $\pm 0.886$ & $\pm 0.505$ &   &\\  
\\
NG-CO       & $2.420$      & $0.07551$    & $0.4242$      &  3.709    & $-2.404$    &$-0.1880$       &  2.098    & $-0.8818$    & 0.1722       & here & 1.94 \\    
           & $\pm 0.029$  & $\pm 0.03588$ & $\pm 0.0083$  & $\pm 0.157$ & $\pm 0.165$ & $\pm 0.0546$ & $\pm 0.232$ & $\pm 0.2976$ & $\pm 0.1408$    &  & \\  %
\\
			\multicolumn{12}{c}{ Comet C/1997~BA$_{6}$} \\    
NG-std   & $3251$       & $88.35 $     & $-25.26$     &   $< 0$ &  &  & un  &  &  & here   &  -- \\    
         & $\pm 115$   & $\pm 50.90$  & $\pm 11.68$   &  &  &  &  &  &  &    &\\  %
		\\
NG-subsolar     & $8.872$      & $0.2761 $    & $-0.04618$      & un &  &  & un  &  & & here  & -- \\   
	water       & $\pm 0.294$  & $\pm 0.1162$ & $\pm 0.0230$    &  &  &  &  &  & &   &\\  
		\\
	NG-CO       & $3.548$      & $0.2619 $    & $-0.01456$      &  1.445    & $-2.734$    & 0.5933      &  3.236    & $-1.641$    & 0.5741       & here  & 0.857 \\    
		        & $\pm 0.139$  & $\pm 0.0644$ & $\pm 0.0089$  & $\pm 0.531$ & $\pm 0.426$ & $\pm 0.146$ & $\pm 0.690$ & $\pm 0.684$ & $\pm 0.216$    &   &\\  
		\\
		\multicolumn{12}{c}{ Comet C/2015~O1} \\    
NG-std   & $13081$       & $-3506 $     & $1065$      & $81427$    & $6476$    & $3631$    & $13517$    & $7145$   & $3703$   & here & 5.20  \\    
         & $\pm 106$   & $\pm 85  $  & $\pm 36$       & $\pm 5645$ & $\pm 895$ & $\pm 567$ & $\pm 2144$ &$\pm 233$ &$\pm 669$ &      &\\  %
\\
NG-subsolar & $9.275$     & $-3.532 $   & $0.7162$     & $24.69$     & $-1.949$    & $-0.0362$    & $4.778 $     &$-0.2231$    &$2.048$     & here  & 4.76\\ 
water       & $\pm 0.069$ & $\pm 0.073$ & $\pm 0.0208$ & $\pm 0.906$ & $\pm 0.412$ & $\pm 0.1248$ & $\pm 0.607 $ &$\pm 0.7771$ &$\pm 0.193$ &       &     \\  %
\\
NG-CO       & $3.200$      & $-1.417$    & $0.2351$       & $3.820$      & $-0.0332$    & $-0.1175$      &  1.098    & $0.2830$      & 0.2711        & here  & 3.28 \\    
            & $\pm 0.025$  & $\pm 0.035$ & $\pm 0.0072$   & $\pm 0.145$  & $\pm 0.1183$ & $\pm 0.0327$  & $\pm 0.187$ & $\pm 0.1754$ & $\pm 0.0421$  &   &\\  
\hline 		
	\end{tabular}
\end{table*}

	\begin{table*}
	\caption{\label{tab:NG_parameters_LAn} NG~parameters $A_1$,
          $A_2$, and $A_3$ (in units of
          $10^{-8}$\,au$\cdot$day$^{-2}$) for the selected NG~orbits
          described in Sec.~\ref{subsec:2000sv74} for C/2000~SV$_{74}$
          from the sample LAn; uncertainties in the NG~parameters are
          listed on the lines below each set of values. Differences in
          the NG~parameters between the CODE Catalog and this study
          are only due to a different approach to data weighting; for
          more discussion, see text.  The last column gives the
          asymmetry in NG parameters between the PRE and POST orbital
          legs; for more discussion, see text.}
        \setlength{\tabcolsep}{3.5pt} \centering
	\begin{tabular}{lrrrrrrrrrrr}
		\hline 	\hline 
		Type of        & \multicolumn{3}{c}{f u l l ~~~a r c} & \multicolumn{3}{c}{P R E} & \multicolumn{3}{c}{P O S T} & References & Asymm.\\ 
		NG~model       & A$_1$   & A$_2$   & A$_3$   & A$_1$   & A$_2$   & A$_3$     &  A$_1$   & A$_2$   & A$_3$       & & PRE/POST\\ 
		\hline 
		\\
		NG-CO   & $3.764$      & $0.1116 $    & $-0.2619$      &  &  &  &  &  &  & CODE    & -- \\    
		& $\pm 0.046$  & $\pm 0.0501$ & $\pm 0.0114$   &  &  &  &  &  &  & Catalog  &\\  
		\\
		NG-CO   & $3.665$      & $0.01162$    & $-0.2184$      &  9.860 & 1.995 & 1.715          & 4.523 & 2.467 & 0.9624   & here   & 1.95\\    
		& $\pm 0.044$  & $\pm 0.04629$& $\pm 0.0106$ & $\pm 0.673$ & $\pm 0.304$ & $\pm 0.173$ & $\pm 0.221$ & $\pm 0.053$ & $\pm 0.068$    &   &\\  
		\\
		NG-subsolar  & $9.294$      & $0.4777$     & $-0.6734$      & 38.468  & 6.1615  & 6.2629 & 11.628 & 9.6056 & 3.7146 & here   & 2.54 \\  
		water        & $\pm 0.110$  & $\pm 0.0982$ & $\pm 0.0285$   & $\pm 2.917$ & $\pm 0.9939$ & $\pm 0.6836$ & $\pm 0.754$ & $\pm 1.0117$ & $\pm 0.3069$  &   &\\  \\ 
        NG-std   & $6065$       & $550.6 $    & $-596.0$     &  $63783$   & $4573$     & $7501$     & $6182$     &  $9431$     &  $2904$    & here    & 5.53 \\    
                 & $\pm 46$     & $\pm 59.7$  & $\pm 21.7$   &  $\pm 9874$& $\pm 1302$ & $\pm 1618$ & $\pm 2765$ &  $\pm 2404$ &  $\pm 665$ &         & \\  %
		\hline 		\end{tabular}
\end{table*}

Solutions discussed here for C/2015~O1 are based on a data arc
spanning 4.8\,yr in the heliocentric distance range of 8.65\,au --
3.73\,au (perihelion) -- 7.28\,au (2020 March 20); pre-discovery
measurements extend the data arc by almost two months (see
Table~\ref{tab:comet_list}). The comet is still observable; at the
beginning of 2023, it is 13.4~au from the Sun, moving outward about
2~au/yr.

As with C/2000~SV$_{74}$, orbit fits for C/2015~O1 show a significant
decrease in RMS for three types of NG~orbits (see
Table~\ref{tab:NG_solutions}) and a much better residual distribution
in the [O-C]-diagram compared to the GR~solution.  However, the
standard NG~orbit gives a slightly worse fit to data than the two
other types of NG~formulas for all three data arcs (full, PRE and
POST); see Table~\ref{tab:NG_solutions}.

Table~\ref{tab:NG_solutions} presents two sets of solutions for this
comet that differ in the data treatments. The first row contains the
results based on data selected and weighted during orbit
determination, and the second row represents the same data, but with
all points equally weighted.  This is a common example of how data
treatment affects the fitting of various NG models to the
data. Regardless of whether the data were weighted or not, it appears
that the NG~solution based on the HJ17 expression gives the best
solution for the entire data arc. However, the [O-C]-diagram for this
NG~solution has small trends in residuals like those of the
NG~solution presented in Fig.~\ref{fig:c2000sv_nakano} (right column,
middle panel). Therefore, neither these tiny differences in the RMS
nor the differences in the [O-C] diagrams allow for a clear choice
between the subsolar water sublimation (HJ17) and CO-driven
$g(r)$-like formulas. The NG~parameters for the chosen NG~solutions
are given in Table~\ref{tab:NG_parameters_LAn_97ba_15o1}.

In addition, Fig.~\ref{fig:c2000sv_nakano} shows that two different
CO-driven formulas (here using weighted data and in the Nakano Notes)
give similar quality fits, and the apparent differences may be mainly
due to the use of different selection and data weighting procedures in
the Nakano Notes and here. 

The GR orbits (weighted data) determined using PRE and POST data arcs
independently give RMS values of 0\farcs 36 and 0\farcs 54 for similar
numbers of observations before and after perihelion (1904 and 2260,
respectively).  At first sight, this suggests a perfectly suitable GR
orbit for the data arc before perihelion compared to the GR orbit for
POST data. However, the [O-C] diagram reveals some tendencies in the
residuals in the GR orbit using PRE data mainly for the first two
oppositions, that is, for heliocentric distances $> 5.5$\,au. A GR
orbit based solely on data in this heliocentric distance range gives
an [O-C] diagram with no visible trends in the data.  However, the
NG~orbit using the CO-driven formula (or subsolar formula of HJ17) and
the entire PRE data gives an RMS drop of only 0\farcs 01 (weighted
data) and no trends in the [O-C]-diagram.

In the case of the POST data arc, we obtained an NG~orbit with
satisfactorily well-determined NG~parameters $A_1$, $A_2$, and $A_3$
(the RMS decreased by 0\farcs 01). However, there is no distinct
preference of the residual distribution in the [O-C]-diagram for the
NG~orbit compared to the GR~orbit.

The original $1/a$ values of all orbits obtained in the study for
C/2015~O1 (and those discussed below) were between 15\,au$_{-6}$ and
55\,au$_{-6}$, and the uncertainty of $1/a_{\rm ori}$ in each orbit
was small, as for C/2000~SV$_{74}$ (less than 2\,au$_{-6}$). It is
worth noting the different $1/a_{\rm ori}$ values obtained for all
three NG solutions.  However, the orbital solutions based on the
CO-driven formula give $1/a_{\rm ori}$ values ranging from
15\,au$_{-6}$ to 35\,au$_{-6}$, depending on the data arc used. The
Nakano value of 22\,au$_{-6}$ is well inside this range. The $1/a_{\rm
  ori}$ value of $(34\pm 2)$\,au$_{-6}$ was obtained for the GR
solution using part of the PRE data (heliocentric distances greater
than 5.5\,au).  All this leads to the conclusion that the preferred
orbits for determining the origin of this comet are either the GR
orbit using part of the PRE data arc (mentioned above) or NG orbits
based on the PRE data arc (sublimation of water ice from the subsolar
point or CO-driven sublimation). The original $1/a$ values of these
three orbits are between 28\,au$_{-6}$ and 38\,au$_{-6}$.

For this comet, we also determined the four NG~parameters for the
asymmetric $g(r(t-\tau))$-like function, where $\tau$ is the time
shift of maximum $g(r)$ relative to perihelion passage.  We found that
this shift (i.e., the time of peak NG forces) is about 80--110 days
before perihelion, depending on the assumed $g(r)$-like form. The RMS
decrease is at the level of 0\farcs02. The direction of this shift is
in line with significantly larger NG~parameters for PRE data than for
POST data, see Table~\ref{tab:NG_parameters_LAn_97ba_15o1}. Similarly,
for C/1997~BA$_6$, we also were able to obtain NG~orbits for both
orbital legs, as well as the NG~model with a $\tau$ shift. In this
case, the time shift was {\em after} perihelion by about
130--140\,days for all three NG~models of motion. This shift is also
consistent with smaller NG parameter values for PRE than POST data,
see Table~\ref{tab:NG_parameters_LAn_97ba_15o1}.

We also tested the CO formula with $r_0=50$\,au and found no
differences in RMS and [O-C] compared to $r_0=10$\,au, but the $A_2$
parameters for PRE and POST data are small, with values comparable to
their uncertainties. Thus, only NG~parameters for the entire data arc
are well-determined. However, this last solution results in a negative
value of the original $1/a$ (similarly as for the PRE solution based
on the same $g(r)$ for C/2014~M4); see Sec.~\ref{sec:original-a}.

\section{Sample LBn, Comets with New Fits for Non-Gravitational~Effects}\label{sec:five_comets}

Comets in this sample have perihelia ranging from 4.5\,au to 8.4\,au
from the Sun. Thus, their NG~orbits were determined solely based on
CO-driven sublimation.

Below, we discuss the five comets with NG~orbits determined in this
study (marked by asterisks in column~[1] in
Table~\ref{tab:comet_list}). All except C/2009~F4 have pre-discovery
measurements.

The Minor Planet Center and the Nakano Notes only offer GR~orbits for
these five comets, whereas the JPL~Small-Body Database Browser
presents an NG~orbit for one of these objects (for C/2015~H2
Pan-STARRS, as of July~2023).

For all comets from this sample, we also tested the CO formula with
$r_0=50$\,au, see Table~\ref{tab:NG_solutions}. However, in this
section, we decided to discuss only NG~solutions with
$r_0=10$\,au. The reason is that, as we indicated in the previous
section for three comets (C/2017~M4, C/2000~SV$_{74}$, and C/2015~O1),
this form of $g(r)$ with $r_0=50$\,au often gives less reliable
solutions, in particular, the transverse NG~parameter can be much
smaller than for $r_0=10$\,au and poorly determined (with
uncertainties as large as the values). However, it is worth noting
that the CO type of $g(r)$ with $r_0=50$\,au typically results in much
smaller values of original $1/a$; see also Sec.~\ref{sec:original-a}.

	\begin{table*}
	  \caption{\label{tab:NG_parameters}
            NG~parameters $A_1$, $A_2$, and $A_3$ (in units of
            $10^{-8}$\,au$\cdot$day$^{-2}$) for the NG~orbits
            described in Table~\ref{tab:NG_solutions} for five comets
            from the sample LBn (CO sublimation) for which the
            NG~effects are not given in the CODE Catalog;
            uncertainties in the NG~parameters are listed on the lines
            below each set of values.}

	\setlength{\tabcolsep}{5.0pt} 
	\centering
	\begin{tabular}{lrrrrrrrrr}
		\hline 	\hline 
		Comet       & \multicolumn{3}{c}{f u l l ~~~a r c} & \multicolumn{3}{c}{P R E} & \multicolumn{3}{c}{P O S T} \\   
		& A$_1$   & A$_2$   & A$_3$   & A$_1$   & A$_2$   & A$_3$     &  A$_1$   & A$_2$   & A$_3$       \\   
		\hline 
		\\
		C/2015 H2         &  $2.818$      & $-0.2128$    & $0.6251$     & &&& &&   \\
		&  $\pm 0.1138$ & $\pm 0.1683$ & $\pm 0.0239$ & &&& &&   \\
		C/2013 V4         & $1.460$       & $-0.3311$    &  0.0         & &&& &&   \\
		& $\pm  0.055$  & $\pm 0.0814$ &              & &&& &&   \\
		C/2009 F4         & $2.097$       & $-0.4568$     & $-0.2141$    & $3.630    $ & $0.2917 $    & $-0.279$     & &&   \\ 
		& $\pm  0.098$  & $\pm 0.0945$ & $\pm 0.0132$  & $\pm 0.486$ & $\pm 0.2417$ & $\pm 0.084$ & &&   \\
		C/2005 L3         & $0.9417$      & $ 0.3879$    & $0.0228$     & &&& $1.906 $    & $-0.3208 $   & 0.0             \\
		& $\pm 0.0502$  & $\pm 0.0705$ & $\pm 0.0118$ & &&& $\pm 0.210$ & $\pm 0.2085$ &               \\
		C/2010 U3         & $0.6273$      & $ 1.897 $    & 0.0  & $ 2.225 $   &$ 6.014 $     &$ 1.362 $& && \\
		& $\pm 0.3554$  & $ \pm 0.360$ &      & $\pm 1.070 $& $\pm 1.000 $ & $\pm 0.169 $ & && \\
		\hline 		\end{tabular}
\end{table*}

\subsection{C/2015~H2 (Pan-STARRS)}

Comet C/2015~H2 was discovered on 2015 April~24, when it was 6.28\,au
from the Sun and about 1.4\,yr before its perihelion passage.
Pre-discovery images found for this comet expand the data arc by
11~months. Thus, the heliocentric distance range covered by data is
7.96\,au -- 4.97\,au (perihelion) -- 9.52\,au during a period of
5.4\,yr; see Table~\ref{tab:comet_list}.

The CODE~Catalog currently shows a GR~orbit based on the data arc
until 2017 November~22. The data arc is now about two years longer, as
the comet was observed until 2019 October~22. The MPC presents a
GR~orbit based on data from 2014 May~19 to 2019 September~3, while JPL
offers an NG orbit using the same data arc we use (as of July 2023).

This is a unique example that demonstrates very strong evidence of
NG~effects in the comet motion in the LBn sample.  However, the
analysis of NG~effects shows atypical behavior along the orbit, as we
describe below.

\begin{figure*}
	\centering
	\includegraphics[width=1.00\columnwidth]{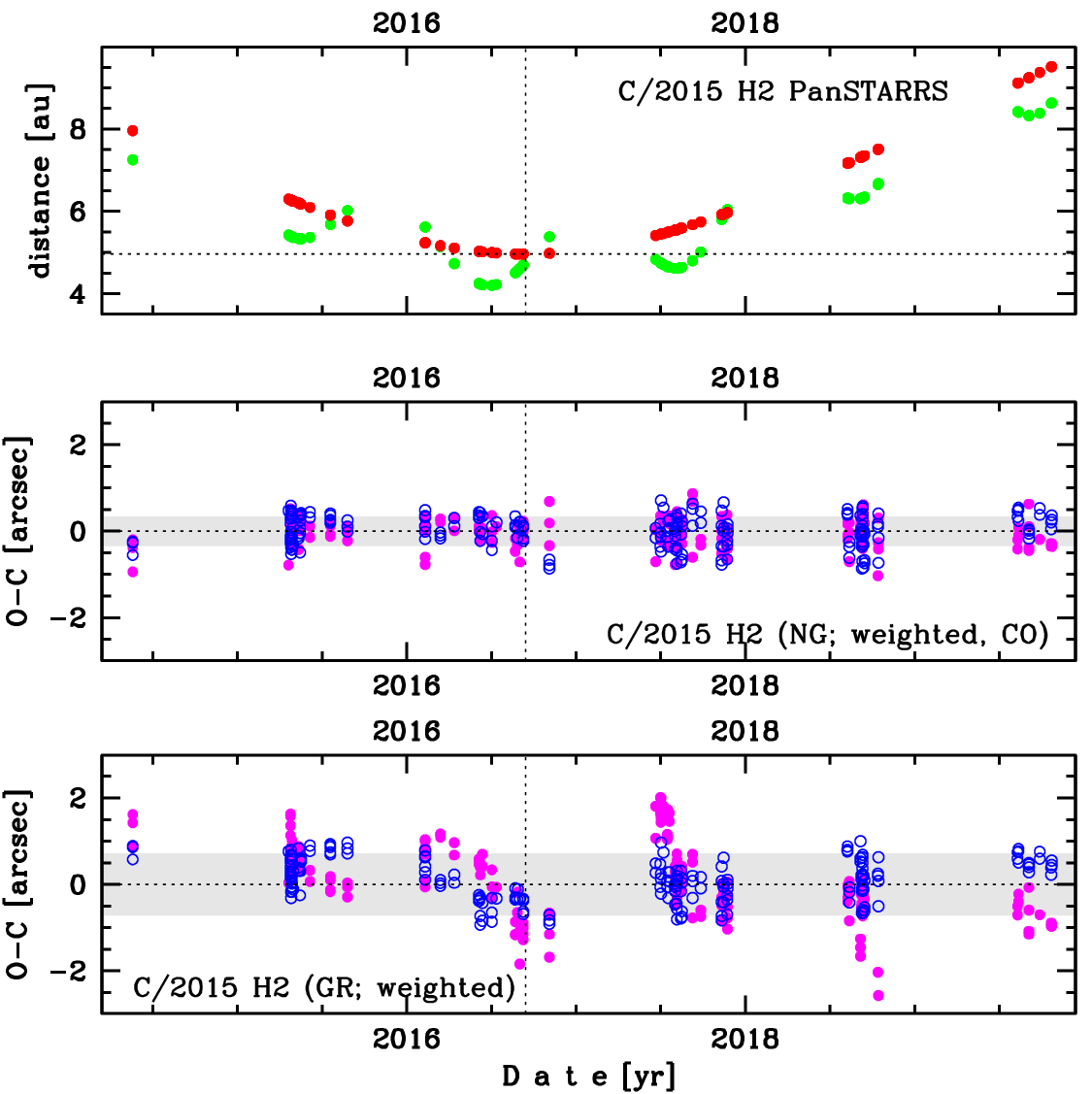}
	\includegraphics[width=1.00\columnwidth]{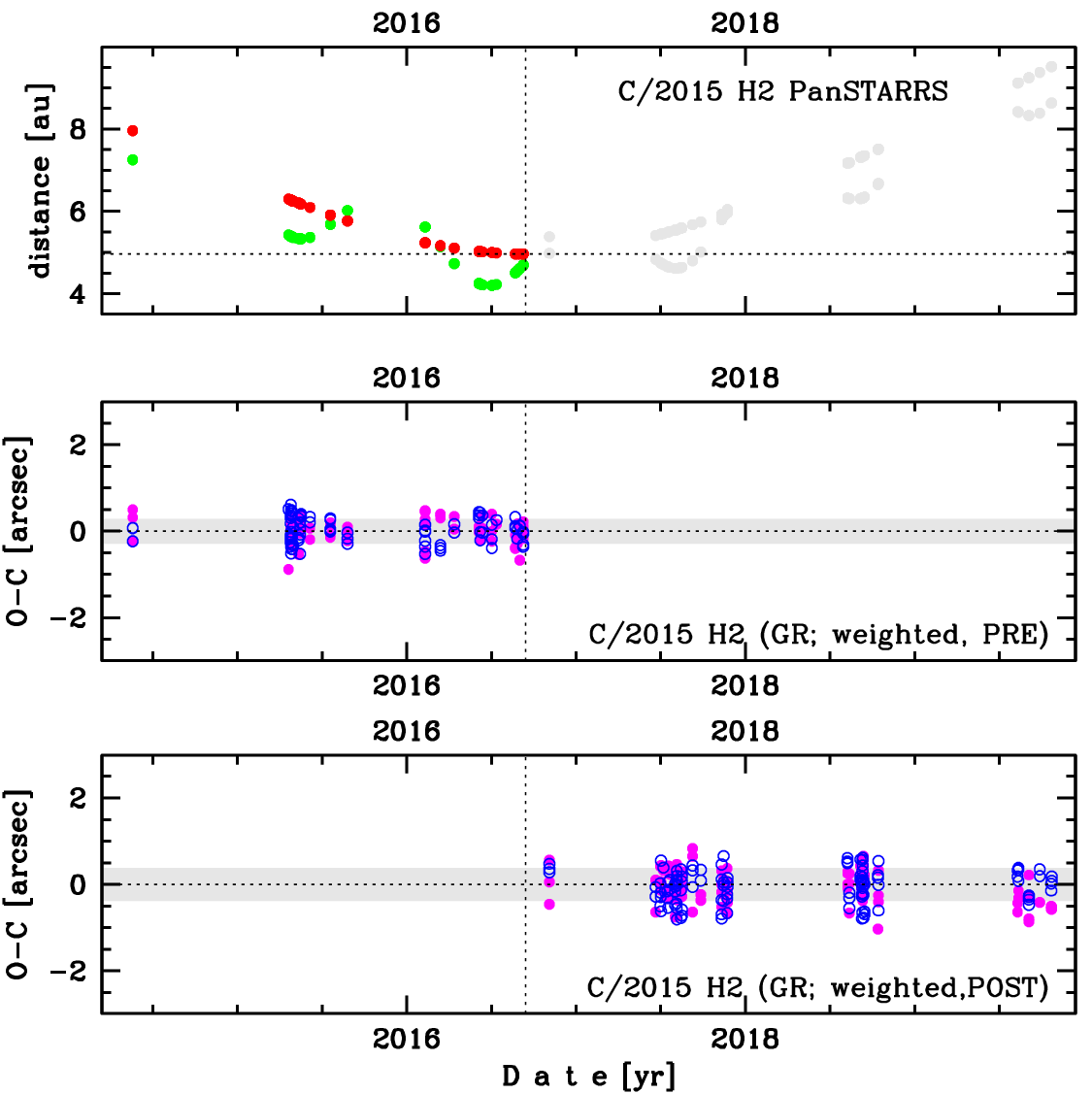}
        
	\caption{Comet C/2015~H2 Pan-STARRS. The left column shows the
          [O-C] diagrams for GR and NG~solutions based on the full
          data arc, while the right column displays [O-C] diagrams for
          two different GR~solutions based on pre-perihelion and
          post-perihelion data arcs, respectively.}
	\label{fig:c2015h2}
\end{figure*}

First, we were able to obtain well-fitted GR~solutions with no firm
evidence of trends in residuals for the PRE and POST data arcs
separately (see Table~\ref{tab:NG_solutions} and the right column of
Fig.~\ref{fig:c2015h2}). In addition, we obtained NG orbits for the
PRE and POST data arcs, but the NG~parameters are uncertain, and the
reductions of RMS are tiny. Therefore, these orbits are not
presented in Tables~\ref{tab:NG_solutions} and~\ref{tab:NG_parameters}.

Second, the GR~orbit based on the complete data arc exhibits notable
residual trends, as we show in the bottom left panel in
Fig.~\ref{fig:c2015h2}. Moreover, the NG~orbit based on the complete
data arc gives a spectacular decrease of RMS from 0\farcs 72 to
0\farcs 34 with no trends in residuals and well-determined radial and
normal components, but a worse-defined transverse component of NG
acceleration (Tables~\ref{tab:NG_solutions} and
\ref{tab:NG_parameters}, and the middle left panel in
Fig.~\ref{fig:c2015h2}). It is the largest relative RMS reduction of
the orbital fits for comets in the LBn sample.

GR and NG~orbits based on the full data arc give values of the
original $1/a$ of $22.6\pm 1.0$~au$_{-6}$ and $34.6\pm 2.0$~au$_{-6}$,
respectively, whereas the GR~orbit based on the PRE data arc gives a
much larger value of $46.4\pm 1.3$\,au$_{-6}$. A very similar tendency
to find smaller values of $1/a_{\rm ori}$ as the comet's data arc
becomes longer can be seen in all the solutions given by the MPC;
note, however, that they provide only GR solutions. JPL presents an
NG~orbit based on the entire data arc and using a CO-driven function
with $r_0=5$\,au for which the transverse component is determined with
even worse accuracy than here. This is probably a consequence of the
smaller value of $r_0$ assumed in the JPL fit (see
Sec.~\ref{subsec:NG_model}).

\subsection{C/2013~V4 (Catalina)}

Comet C/2013~V4 was discovered on 2013 November~9, less than a year
before its perihelion. Pre-discovery images expand the data arc by
about 16~days. This comet was observed for 4.8\,yr until 2018 August~6
in the heliocentric distance range of 7.32\,au -- 5.19\,au
(perihelion) -- 8.99\,au; see Table~\ref{tab:comet_list}.

The CODE Catalog offers only a GR orbit based on the entire data arc
as of July 2023, and this orbit is used here. However, the NG orbit
based on the complete data arc leads to a satisfactory pattern of
residuals in the [O-C]-diagram, where almost all trends disappear, and
the RMS is reduced by about 0\farcs 03 compared to the GR~orbit
(Table~\ref{tab:NG_solutions}).  This NG~solution gives
well-determined radial and transverse components of the
NG~accelerations by assuming a normal component of zero
(Table~\ref{tab:NG_parameters}). Likewise, there are no trends in GR
orbits obtained independently for the PRE and POST data arcs. We were
not able to obtain NG orbits for the PRE and POST data arcs.

GR and NG orbits based on the entire data arc give consistent values
of the original $1/a$ of $79.2\pm 0.2$~au$_{-6}$ and $79.6\pm
0.8$~au$_{-6}$, respectively, whereas the GR orbit using the PRE data
arc gives a slightly larger value ($84.1\pm 0.5$\,au$_{-6}$).

Values of 79.4\,au$_{-6}$ and 68\,au$_{-6}$ for the original $1/a$ are
given by the MPC and the Nakano Notes (NK\,4294) for GR~orbits,
respectively.

\subsection{C/2009~F4 (McNaught)}

\begin{figure*}
	\centering
	\includegraphics[width=1.00\columnwidth]{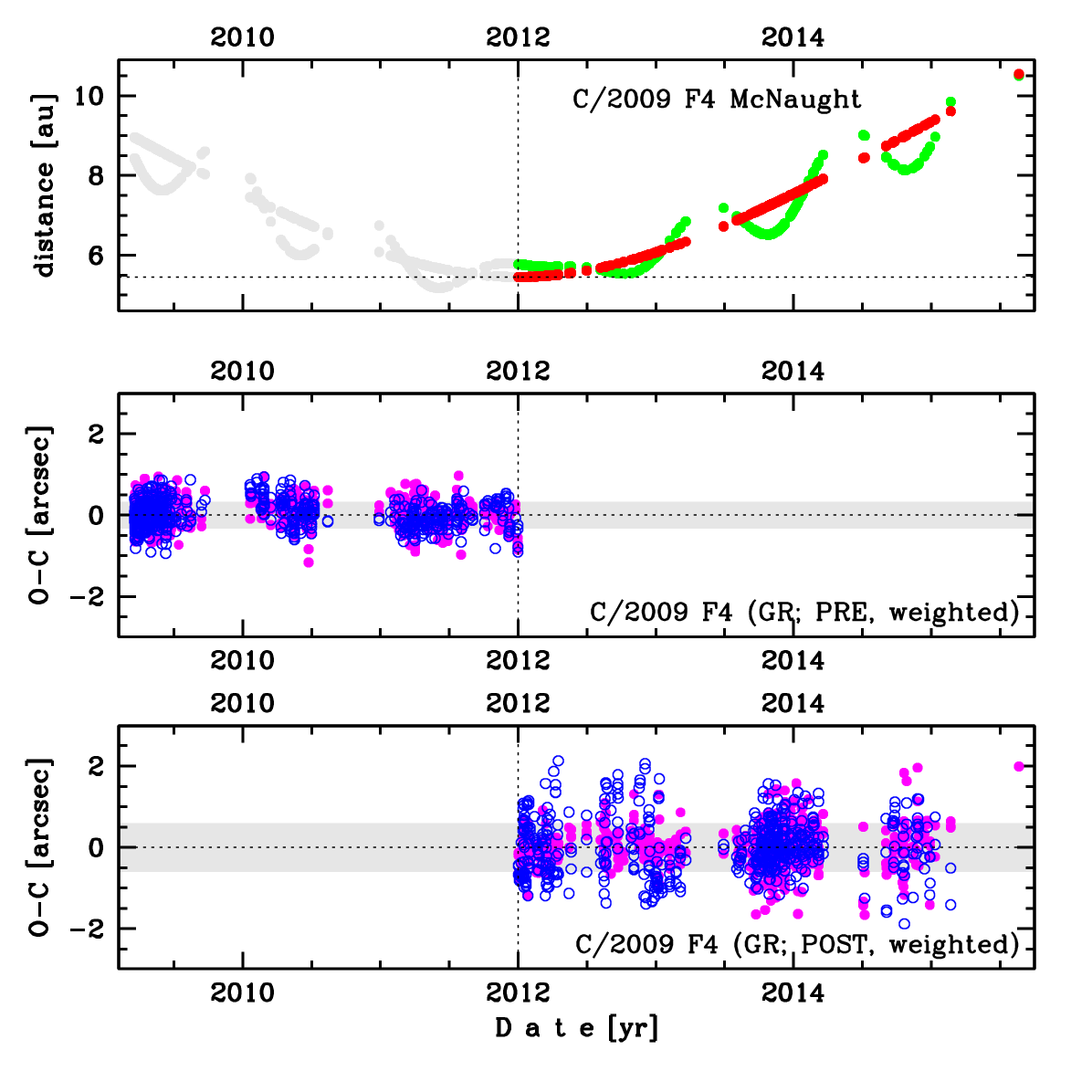}
	\includegraphics[width=1.00\columnwidth]{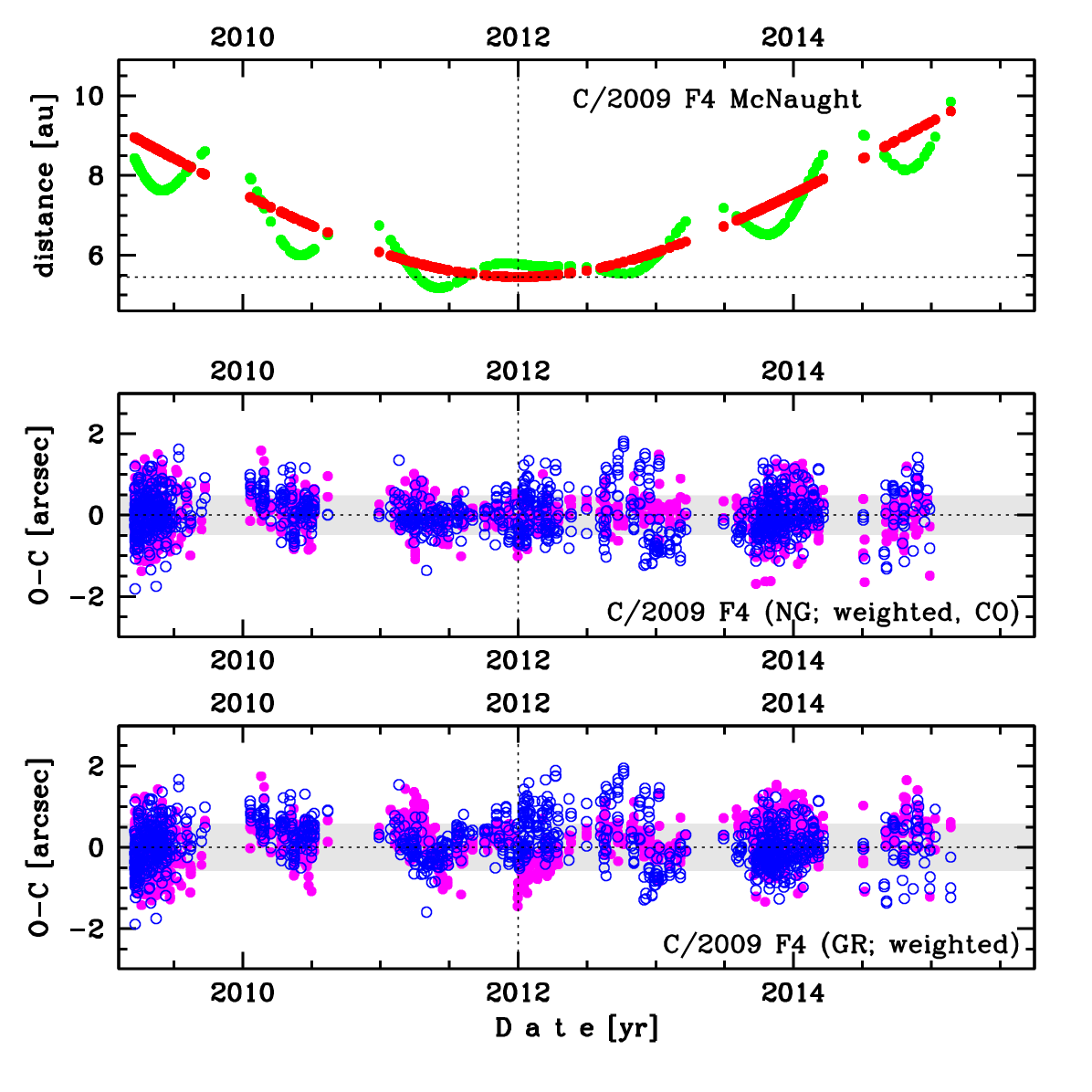}
        
	\caption{Comet C/2009~F4 McNaught. Upper panels: Time
          distribution of positional observations. Lower panels: O-C
          diagrams for selected orbital solutions described in the
          text; for the meaning of the lines and symbols, see
          Fig.~\ref{fig:c2000sv_nakano}.}
	\label{fig:c2009f4}
\end{figure*}

Comet C/2009~F4 was observed for 6.4\,yr during seven oppositions in
the heliocentric distance range of 8.96\,au -- 5.45\,au (perihelion)
-- 10.55\,au; see Table~\ref{tab:comet_list}.

At the end of the data arc, there is a 6-month break with no
observations, and then only two measurements (on 2015 August 20) are
available. Of the last two, only one measurement in right ascension
can be fit well to the orbit determined using the POST data
arc. However, both measurements are completely rejected in the GR and
NG orbits using the complete data arc (the GR~orbit is only given in
the CODE Catalog as of July~2023).  Because of this issue with the
final two data points, we list two sets of data for this comet in
Table~\ref{tab:comet_list}.

The right column of Fig.~\ref{fig:c2009f4} shows the difference
between the data fit to the GR orbit (bottom panel) and the NG orbit
based on the $g(r)$~function for CO sublimation.  Here, the last two
observations are rejected. Thus, both orbits were obtained using a
half-year shorter data arc (see the second row in
Table~\ref{tab:comet_list} for C/2009 F4). The RMS reduction of the NG
orbit compared to the GR orbit is more than 10\% (from 0\farcs 54 to
0\farcs 48) and the NG parameters are formally well-defined
(Table~\ref{tab:NG_parameters}). In addition, the time trends in the
[O-C] diagram are smaller, particularly around perihelion from early
2011 to mid-2012. JPL and Nakano provide GR~orbits using the same data
arc through February 2015, while the MPC orbits are also GR-type and
are given both with and without the August 2015 observations.

Orbits obtained separately using the PRE and POST data arcs provide
interesting results. Fits to both GR~solutions (PRE and POST) to the
respective data are shown in the left column in
Fig.~\ref{fig:c2009f4}.  The much larger residual scatter after
perihelion can be due to several factors, including an increase in
comet activity from various regions of the rotating nucleus. We
describe a similar RMS increase after perihelion for comet C/2015~O1
in Sec.~\ref{subsec:2015o1}.
 
The GR~orbit fitted to the PRE data (618 observations) gives an RMS of
0\farcs 33. This GR~solution already gives a much better fit than is
obtained for NG~solutions based on a full data arc; compare the middle
panel of the left column in Fig.~\ref{fig:c2009f4} to the middle panel
of the right column. In addition, we obtain an NG~orbit with
well-determined uncertainties for the radial and normal components of
the NG acceleration, but the transverse component is poorly
determined; the RMS decreases by only $\sim$0\farcs 01, but the time
trends in the [O-C] diagram are reduced.

The situation is quite different for the POST data arc. Here, the RMS
of the GR~orbit is 0\farcs 60 (595 observations; the two August 2015
measurements are almost completely rejected), which is twice as large
as for the PRE data arc.  Furthermore, it is impossible to reasonably
determine an NG orbit with the POST data arc, even if both
observations from August~2015 are completely rejected.

GR and NG~orbits based on the complete data arc through February 2015
result in original $1/a$ values of $40.6\pm 0.3$ and $44.3\pm 0.7$
au$_{-6}$, respectively, whereas GR and NG~orbits based on the PRE
data arc give slightly larger $1/a_{\rm ori}$ values ($49.9\pm
0.5$\,au$_{-6}$ and $46.3\pm 1.0$\,au$_{-6}$). The PRE orbits seem
best suited for studying the origin of this comet. The GR~solutions in
the MPC and Nakano Notes give 46.3 and 31 au$_{-6}$ for $1/a_{\rm
  ori}$, respectively.

\subsection{C/2005~L3 (McNaught)}

Solutions obtained in this study are based on data spanning 9.5\,yr in
the heliocentric distance range of 10.3\,au -- 5.59\,au (perihelion)
-- 14.93\,au; two series of four pre-discovery detections extend the
data arc by about 10.5~months (see Table~\ref{tab:comet_list}).  The
CODE~Catalog gives only a GR~orbit as of July~2023, using a data arc
more than a year shorter than that available now. Now, NG~effects are
quite well-determined using the longer data arc and a $g(r)$-like
function reflecting CO sublimation; however, the RMS decreases only by
0\farcs 01 compared to the GR~orbit (see
Tables~\ref{tab:NG_solutions}~and~\ref{tab:NG_parameters}). No
reliable NG~orbit can be obtained based on the 3.4-year PRE data arc.

GR orbits based on the full data arc and PRE data give consistent
values of the original $1/a$ of about $61.7\pm 0.1$ and $62.1\pm 0.6$
au$_{-6}$, respectively, whereas the NG~orbit based on the full data
arc gives a slightly larger value ($68.4\pm 0.7$\,au$_{-6}$).  The
Minor Planet Center and the Nakano Notes list GR orbits with $1/a_{\rm
  ori}$ of 61.25\,au$_{-6}$ and 51\,au$_{-6}$, respectively.

\begin{figure*}
	\centering
	\includegraphics[width=1.00\columnwidth]{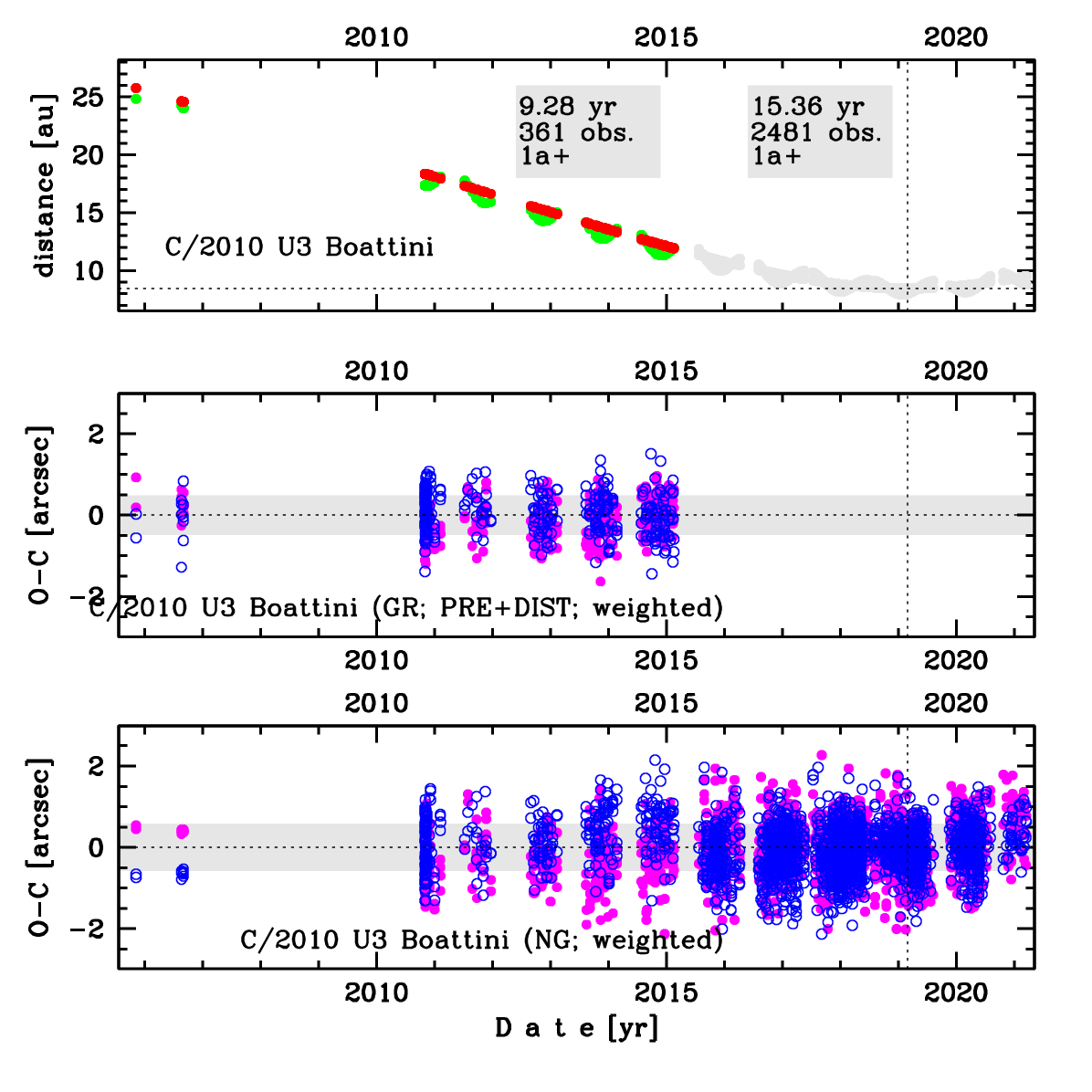}
	\includegraphics[width=1.00\columnwidth]{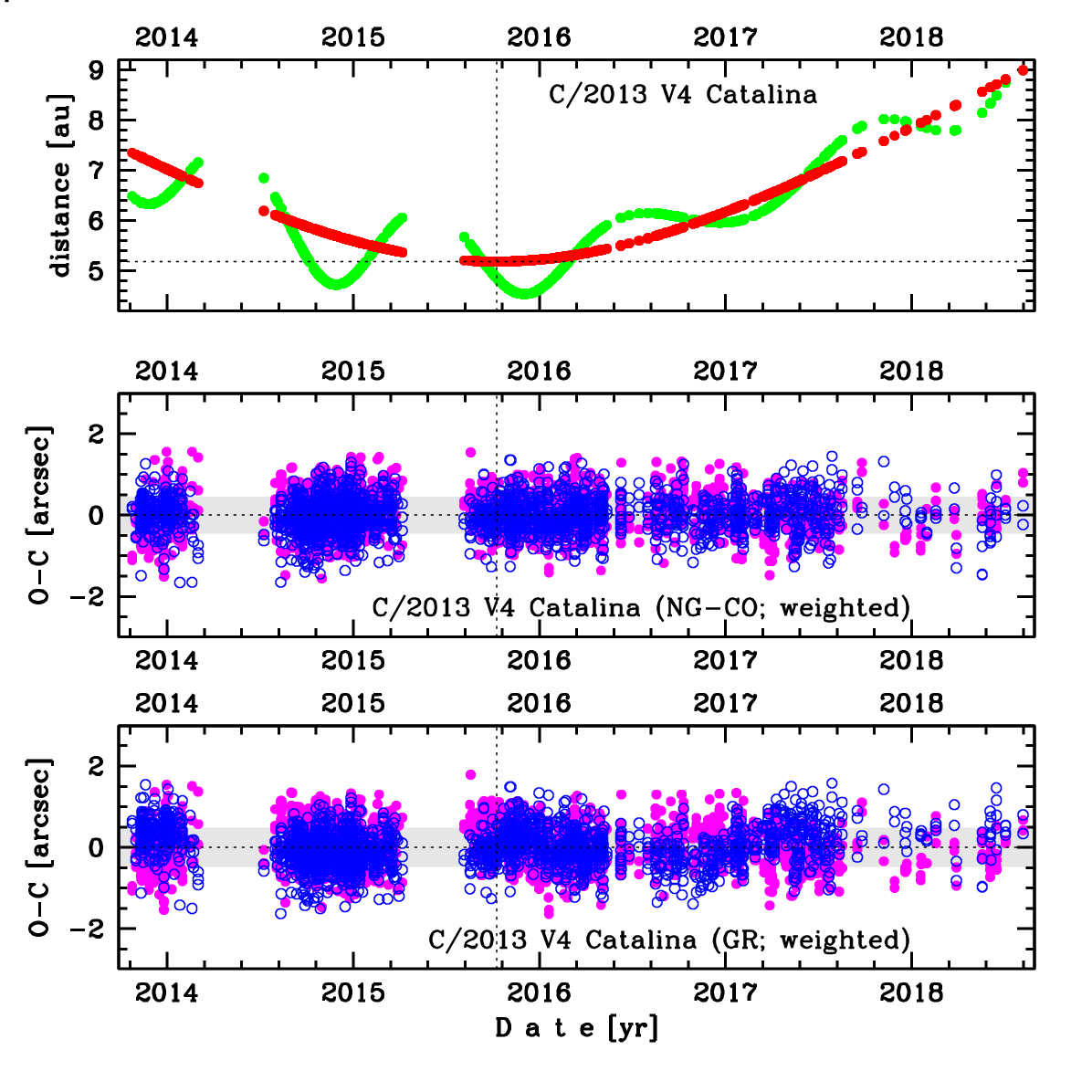}
        \caption{Comets C/2010~U3 Boattini (left column) and C/2013~V4
          Catalina (right column). Upper panel: Time distribution of
          positional observations. Lower panels: O-C diagrams for
          selected orbital solutions described in the text; for the
          meaning of the lines and symbols, see
          Fig.~\ref{fig:c2000sv_nakano}.}
	\label{fig:c2010u3}
\end{figure*}

\subsection{C/2010~U3 (Boattini)}\label{subsec:2010u3}

C/2010~U3 was discovered on 2010 October~31, but several pre-discovery
images extended the data arc back to 2005 November~5, when the comet
was 27.75\,au from the Sun. At the moment of discovery, it was more
than 8\,yr before its perihelion passage and 18.4\,au from the
Sun. This comet passed perihelion on 2019 February~26 at a distance of
8.45\,au from the Sun and was still observed in April
2023\footnote{\tt \tiny
https://minorplanetcenter.net/iau/lists/LastCometObs.html}.
\citet{Hui:2019a} indicate that this comet was active in late 2005
when it was well outside Uranus's orbit (at $\sim$26\,au from the
Sun).  They examined the pre-perihelion data arc from November 2005 to
December 2018; the authors found many of the pre-discovery
measurements by using the Solar System Object Image Search service
\citep{Gwyn2012}\footnote{ {\tt \tiny
  https://www.cadc-ccda.hia-iha.nrc-cnrc.gc.ca/en/ssois/}}.  Assuming
a pure GR~model of motion, \citet{Hui:2019a} concluded that over half
the measurements (984 of the total of 1723 positional observations,
RMS = 0\farcs 41) should be rejected due to poor internal consistency.
Using their GR~orbit, \citet{Hui:2019a} studied its dynamical
evolution and calculated that the comet passed its previous perihelion
at a distance of $8.364\pm 0.004$\,au from the Sun approximately
1.96\,Myr ago. Thus, they concluded that this comet was a dynamically
old member of the Oort Cloud. Their model includes galactic tides, but
not perturbations by passing stars during the 2-Myr period covered by
their dynamical calculations.  However, some stars may strongly
perturb the motion of Oort Cloud comets during this time. Some
examples of the strongest stellar perturbers are given in
\citet{Dyb-Bre:2021}, \citet{dyb-kroli:2022}, and
\citet{Bailer-Jones2022}. For this reason, the estimation that
C/2010~U3 is a dynamically old comet by \citet{Hui:2019a} requires
recalculation with the new set of stars mentioned in the articles
cited above.

\citet{Hui:2019a} also attempted to determine an NG~orbit using the
same data set described above and found that the NG~orbit is defined
marginally (the uncertainties of the three NG~parameters are
comparable to their values, see Table~3 therein) and did not improve
the orbital fits.

The CODE Catalog currently (as of April 2023) includes two GR~orbital
solutions with their [O-C]-diagrams.  The first orbit is based mainly
on pre-perihelion data spanning 14.25\,yr in the heliocentric distance
range of 25.75\,au -- 8.45\,au (perihelion) -- 8.68\,au (2020
February~3, solution `b5', 2144 observations, RMS = 0\farcs 58). This
solution gives systematic deviations in right ascension (positive
values) and declination (negative values) for all 11~pre-discovery
measurements and some tiny trends in part of the data.  The second
orbit (solution `p5') is obtained using part of the PRE data arc and
consists of measurements taken at heliocentric distances greater than
14.88\,au (data until 2013 February~8). This last solution provides a
good fit to the data arc, showing the high quality of the
pre-discovery data set (see solution `p5' in the CODE~Catalog, 206
observations, RMS = 0\farcs 53).

The GR~orbit fit to the data set for the latter case and traces of a
trend in residuals for the GR~orbit obtained using the observations
until about 2\,yr after perihelion passage suggest the existence of
NG~effects in the positional data, despite the comet's large
perihelion distance. Data spanning 15.4\,yr (observations until 2021
April~18) allow us to conclude that an NG~orbit can be marginally
determined using the available data arc (Table~\ref{tab:comet_list}),
with the radial and transverse components of the NG acceleration at
significance levels of $2\sigma$ and $4\sigma$, respectively
(Table~\ref{tab:NG_parameters}). However, the [O-C] diagram for this
NG~orbit based on the full data arc (lowest panel in the left column
in Fig.~\ref{fig:c2010u3}) still shows trends in residuals like those
for the GR~orbit based on the same data arc. The NG~parameters based
only on the PRE data are somewhat more reliable
(Table~\ref{tab:NG_parameters}).  However, the GR~orbit based on the
9.28\,yr data arc until mid-February 2015 (361 observations, RMS =
0\farcs 49) seems to fit the data better than the NG~orbit obtained
using the complete data arc (2481 observations, RMS = 0\farcs 58, see
also Table~\ref{tab:comet_list}), because the shorter arc shows no
trends in residuals (compare the middle panel with the bottom one in
the left column in Fig.~\ref{fig:c2010u3}).

Therefore, we conclude that C/2010~U3 is the comet with the largest
perihelion distance for which NG~accelerations seem to be detectable
using positional measurements in the period 2005 November -- 2021
March (Table~\ref{tab:NG_solutions}).  Additionally, the NG~orbit
obtained with the CO formula shows a trend in the residuals (as in the
case of the GR~orbit), which already occurs in the PRE data at
heliocentric distances between 10 and 11\,au. Unfortunately, using a
larger value of $r_0$ (15, 20, 30 and 50\,au) does not improve the fit
to the data.

\begin{table}
  \caption{\label{tab:2010u3_pre_solution_series} Series of
    GR~solutions obtained for C/2010~U3 using three different sets of
    pre-perihelion data arcs, each starting from 2005
    November~5. Column~[2] gives the comet's heliocentric distance at
    the time given in Column~[1], while Columns~[3]--[5] refer to 11
    pre-discovery measurements taken at Mauna Kea in 2005--2006.}
  
	\centering
	\setlength{\tabcolsep}{2.5pt} 
	\begin{tabular}{ccccccc}
		\hline 			\hline 
		\multicolumn{2}{c}{end of data arc} & relative  & 	\multicolumn{2}{c}{[O-C] trends} & RMS       & 1/a$_{\rm ori}$     \\
		date   & r$_h$    & weights   &  $\alpha$ & $\delta$                     & [arcsec]  & [au$_{-6}$]         \\
		\hline \\
		2014\,02\,22  & 13.34 & 6.01  & no  & no  & 0\farcs 51 & $57.2 \pm 1.3$ \\
		2015\,02\,15  & 11.92 & 5.25  & no  & no  & 0\farcs 49 & $59.1 \pm 1.0$ \\
		2016\,04\,04  & 10.44 & .607  & yes & yes & 0\farcs 58 & $61.1 \pm 1.0$ \\
		\hline
	\end{tabular}
\end{table}

To examine the possible existence of NG~forces and $1/a_{\rm ori}$
changes within the pre-perihelion leg of C/2010~U3's orbit more
carefully, a series of GR~orbits using increasingly longer data arcs
was obtained, starting from the orbit with one opposition more than
that given in the CODE Catalog (solution `p5', data arc until early
2013). Each of the following three GR orbits was obtained using a data
arc with one opposition more than the previous one, i.e., using data
through early 2014, 2015, and 2016; see
Table~\ref{tab:2010u3_pre_solution_series} and the upper left panel in
Fig.~\ref{fig:c2010u3}. Here, eleven precise pre-discovery
observations obtained in 2005 and 2006 at Mauna Kea Observatory are
perfect for finding traces of systematic trends (which we identify
with the existence of NG~acceleration in the comet motion) in this
series of solutions. A tendency towards clearly decreasing relative
weights for these 11 pre-discovery measurements as the data arc
lengthens is easy to see, starting from a level several times better
than the mean weight of observations to about half of this mean
(column [3] of Table~\ref{tab:2010u3_pre_solution_series}; the
weighting procedure is described in detail in
\cite{krolikowska-sit-soltan:2009}). Moreover, for the last orbit of
this series (data until April 2016), a trend in residuals was also
visible: all residuals in right ascension were positive, and in
declination, negative; a similar tendency can be seen for the
NG~solution discussed earlier (bottom left panel in
Fig.~\ref{fig:c2010u3}). From the analysis of this series of
solutions, we conclude that the orbit based on the longest data arc,
not yet showing NG effects, is the one obtained using the data arc
from 2005 November~5 -- 2015 February~15 (middle left panel in
Fig.~\ref{fig:c2010u3}), which included five oppositions after the
discovery of C/2010~U3 in 2010 (middle solution in
Table~\ref{tab:2010u3_pre_solution_series}). It means that tiny traces
of NG~effects start to appear in the positional data when this comet
was beyond 11\,au from the Sun.

All GR and NG~orbits obtained in this study (based on different data
arcs) give very similar values of the original semimajor axes between
55 and 65 au$_{-6}$. However, the preferred starting orbit for
dynamical studies of the comet's past evolution is the orbit based on
the PRE data arc up to 2015 February~15, which has $1/a_{\rm ori} =
59.1 \pm 1.0$~au$_{-6}$.

\section{Outline of a Few Other Cases}\label{sec:other_examples}

This section describes three more comets from the sample LBn --
C/2008~FK$_{75}$ (Lemmon-Siding Spring), C/1999~U4 (Catalina-Skiff),
C/2006~S3 (LONEOS), and the comet with the longest data arc of objects
from the LBg sample, C/2008~S3 (Boattini).  All four comets, except
C/2008~FK$_{75}$, have a series of pre-discovery measurements. MPC,
JPL, and the Nakano Notes present NG~orbits only for
C/2006~S3~(LONEOS).

	\begin{table*}
          \caption{\label{tab:NG_parameters_2} The same as in
            Fig~\ref{tab:NG_parameters} for the remaining comets from
            the Sample~LBn (CO-sublimation).         
      	    Two rows of solutions in the case of C/2008~FK$_{75}$
            correspond to the respective rows of solutions described
            in Table~\ref{tab:NG_solutions}; for further discussion,
            see Sec.~\ref{subsec:2008fk}.
            }
          
	\setlength{\tabcolsep}{3.5pt} 
	\centering
	\begin{tabular}{lrrrrrrrrrrr}
		\hline 	\hline 
		Comet       & \multicolumn{3}{c}{f u l l ~~~a r c} & \multicolumn{3}{c}{P R E} & \multicolumn{3}{c}{P O S T} & References & Asymm \\   
		& A$_1$   & A$_2$   & A$_3$   & A$_1$   & A$_2$   & A$_3$     &  A$_1$   & A$_2$   & A$_3$       &           &PRE/POST \\   
		\hline 
		\\
		C/2008 FK$_{75}$  & $1.861$       & $-0.5112$    & 0.2751     &  &   &   &   &    & & KD17, CODE  & \\    
		& $\pm 0.048$   & $\pm 0.0605$ & $\pm 0.0111$&   &            &  &   &   &   &                    & \\
		& $2.006$       & $-0.4252$    & 0.2971      &  1.322 & $-1.736$ & 0.0774 & 1.980 & 0.3318& $-$0.0765 & here  &  1.18 \\    
		& $\pm 0.031$   & $\pm 0.0426$ & $\pm 0.0083$& $\pm 0.196$    & $\pm 0.141$ & $\pm 0.0663$ & $\pm 0.213$ & $\pm 0.2719$ & $\pm 0.0448$   &  & \\
		C/1999 U4         &  $0.8257$      & $-1.095$    & $0.1733$     & &&& && & here &  \\
		&  $\pm 0.1566$ & $\pm 0.185$  & $\pm 0.0249$ & &&& &&   & & \\
		C/2006 S3         & $1.776$       & $2.098$      & 0.2560       &  1.184 & 0.6103 & 0.3725 & 3.776 & 4.356 & $-$0.8553  & here       & 0.24 \\
		& $\pm 0.033$   & $\pm 0.035$ & $\pm 0.0061$  & $\pm 0.207$ & $\pm 0.2082$ & $\pm 0.037$ & $\pm 0.159$ & $\pm 0.156$ & $\pm 0.0339$  &  & \\
		\hline 		\end{tabular}
\end{table*}

\subsection{C/2008~FK$_{75}$ (Lemmon-Siding Spring)}\label{subsec:2008fk}

The data arc of C/2008~FK$_{75}$ covers 5.6\,yr in the heliocentric
distance range of 8.24\,au -- 4.51\,au (perihelion) -- 9.51\, au. The
CODE Catalog provides an NG~orbit based on the CO-driven $g(r)$-like
function and complete data arc. That orbit gives a better distribution
of residuals in the [O-C] diagram, especially for the first and last
oppositions, than does the GR~orbit.

In this study, we successfully determined NG orbits for the PRE and
POST data arcs independently. However, in the case of the POST data,
only the radial component of the NG~acceleration is well-determined;
see Table~\ref{tab:NG_parameters_2}.  Each of the GR or NG orbits
based on the PRE and POST data yields [O-C] distributions of the same
quality in the respective data arc as the NG~orbit based on full data
arcs (so obtained by assuming three constant NG~parameters within the
full period of observations).

Additionally, we redetermined the GR and NG orbits using the full data
arc applying the data selected and weighted as for GR~orbit
determination using PRE and POST data, independently, and the
residuals of these orbits are given in
Table~\ref{tab:NG_solutions}. There are only tiny differences in
NG~parameters between our previous results used in KD17 (first row in
Table~\ref{tab:NG_solutions} for this comet; also see solutions `a6'
and `c6' in the CODE Catalog) and the NG~orbit obtained here (first
row in the table for this comet).

GR and NG~orbits based on the complete data arc (redetermined here and
those from KD17) result in a narrow range of original $1/a$ between
27.9 and 42.9 au$_{-6}$. The GR orbit obtained using the PRE data
gives a similar original $1/a_{\rm ori} =38.5\pm
0.3$\,au$_{-6}$. However, the NG-CO solution based on PRE data
suggests that this value might be as small as
17.8\,au$_{-6}$. GR~orbital solutions presented in the MPC and the
Nakano Notes give values of 29\,au$_{-6}$ and 18\,au$_{-6}$,
respectively, for $1/a_{\rm ori}$.

\subsection{C/1999~U4 (Catalina-Skiff)}

C/1999~U4 was discovered on 1999 October~31 by the Catalina Sky Survey
and, independently, on November~1 by Brian~A.~Skiff of the LONEOS
Survey. Pre-discovery detections from 1999 September 18 and October 30
were later found.  The CODE~Catalog currently (July~2023) presents GR
and NG orbits based on data spanning 4.6\,yr in the heliocentric
distance range of 7.55\,au -- 4.92 au (perihelion) -- 8.22 au. We use
the same data arc here.

Unfortunately, the NG~orbits for PRE and POST data arcs are uncertain,
and we do not consider them here.  GR and NG~orbits obtained
previously (CODE Catalog) and those calculated here using complete and
PRE data arcs result in original orbits in a narrow range of original
$1/a$ from 22 to 38 \,au$_{-6}$. Each orbit has a small $1/a_{\rm
  ori}$-uncertainty of at most 3\,au$_{-6}$.  GR~orbits in the MPC and
the Nakano Notes give values of 36.9\,au$_{-6}$ and 21\,au$_{-6}$ for
$1/a_{\rm ori}$, respectively.

\begin{figure}
	\centering
	\includegraphics[width=1.00\columnwidth]{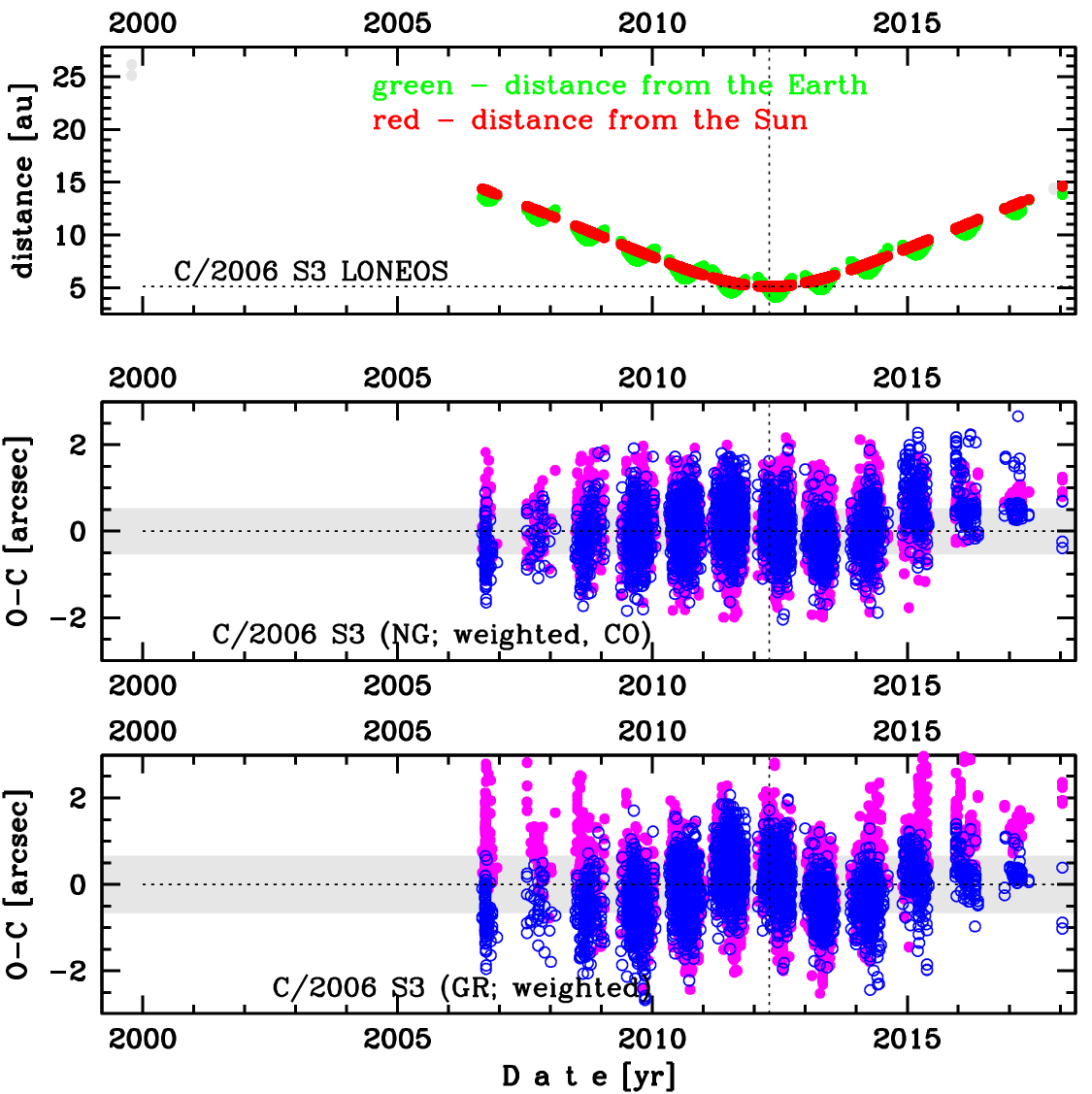}
        	\caption{Comet C/2006~S3 LONEOS. Upper panel: Time
          distribution of positional observations. Lower panels: O-C
          diagram for GR (bottom panel) and NG (middle panel) orbits
          obtained using the full data arc; for the meaning of the
          lines and symbols, see
          Fig.~\ref{fig:c2000sv_nakano}. Pre-discovery data were
          rejected because the residuals were too large.}
        
	\label{fig:c2006s3}
\end{figure}

\subsection{C/2006~S3 (LONEOS)}\label{subsec:2006s3}

Solutions discussed here are based on a data arc spanning 18.3\,yr in
the heliocentric distance range of 26.14\,au -- 5.13\,au (perihelion)
-- 14.64\,au. Our fits include three pre-discovery positional
measurements from 1999 October~13, and two more from 2006~August~29
and September~17. The latter series occurred two days before the comet
was discovered on 2006 September 19. At the time of discovery,
C/2006~S3 was 14.3 au from the Sun. The comet passed perihelion on
2012 April~12.

Photometric and spectroscopic studies of C/2006~S3 by
\citet{shi-ma-zheng:2014_2006s3} and \citet{rouss-etal:2014_2006s3}
showed that this comet was very active at $\approx 6$~au, producing
dust at a rate like that of 29P/Schwassmann–Wachmann~1 (also see
\cite{Kulyk2018}).

When we performed orbit fits for this comet at the end of 2021, the
data available at the MPC extended about 1.6\,yr longer, until 2018
January 16, compared to our previous results (KD17, the CODE
Catalog). As a result, all kinds of orbits for this comet have been
recalculated here. We were able to obtain an NG~orbit using the entire
data arc, as well as for the PRE and POST data arcs; see
Table~\ref{tab:NG_solutions}.

For this comet, two different NG~solutions were presented at the MPC
and at JPL as of July~2023. The MPC's orbit was based on data spanning
from 2006 August~29 through 2020 November~24, and JPL has a
NG~solution that goes from 2006 September 19 through 2020 November~24.
JPL's data arc covers the full period of positional detections since
the comet's discovery, but does not take pre-discovery measurements
into account.

We found that the 1999 measurements (in right ascension and
declination) did not fit the orbit for the entire data arc as well as
for the PRE data arc. Moreover, these 1999 data give large residuals
in declination even for the orbit determined using distant
pre-perihelion (PRE+DIST) data.  We indicate this problem by including
two data arcs of different lengths in Table~\ref{tab:comet_list}.

Using the data from 1999 to 2009 December 19, in the heliocentric
distance range between 26.14\,au and 8.01\,au, the measurements in
right ascension are well-fitted to the NG~orbit, while those in
declination are rejected, and some traces of NG acceleration are
detectable at the level of 2$\sigma$. Furthermore, the measurements in
right ascension are closer to the NG orbit than they are for the GR
orbit. The large residuals in declination do not seem to be improved
by assuming a larger value of $r_0$.

Fig.~\ref{fig:c2006s3} shows that a model with three constant
NG~parameters from 2006 until 2018 is not satisfactory. In particular,
it still shows some wavy trends after perihelion. Introducing a fourth
NG parameter, i.e., a time advance or delay in the asymmetric
$g(r)$-like function, does not improve this situation and does not
reduce the RMS. However, we determined two NG solutions using PRE and
POST data arcs; see Table~\ref{tab:NG_parameters}. The NG~parameters
of both these solutions show that the magnitude of the NG~acceleration
is about 4.2~times greater for the post-perihelion leg of the orbit
than for the pre-perihelion leg.  This result agrees with the
conclusion of \citet{rouss-etal:2014_2006s3} that C/2006~S3 was more
active after its perihelion than before (see Figure~5
therein). Assuming constant NG~parameters after perihelion still shows
some time trends in the orbital fits, although they are smaller than
for a model with constant NG~parameters along the entire data arc. The
comet may have undergone an outburst in February 2014 at 7.16~au
\citep{sarneczky-et-al:2016}, almost two years after perihelion
passage. Even when we limit the data arc to mid-2015, some trends are
still visible in the residuals for the NG~orbit.  Thus, we can
conclude that the comet's activity must have changed after perihelion
in 2012, but before mid-2015.  It is worth noting that the MPC, in
addition to the two NG~orbits (the first mentioned earlier and the
second covering observations up to 2016 November 29), also offers a
set of GR solutions based on shorter and longer datasets, but always
starting with observations from 2006 August 29.

In May 2022 we found that 16 more observations from 2018 May~16 to
2020 November~24 (heliocentric distance range 15.25\,au -- 19.76\,au)
were available at the MPC, so we used these data to validate the above
conclusion. These measurements were taken at Mauna Kea (from 2018 May
16 and 2018 June 18, observatory code 568) and by the Dark Energy
Survey (from 2018 December~12 to 2020 November~24, observatory code
W84). We constructed a series of three different data arcs, where the
first included data from 2015 December~10 to 2020 November~24, and the
second and third included one and two more oppositions than were used
in the first data set, respectively; see
Table~\ref{tab:2006s3_post_solution_series}. We fit GR~orbits to these
three data sets.  For the first two data sets, we obtained very good
GR~orbit fits with no trends in relative weights and [O-C] time
distribution, and no traces of NG~effects. However, for the longest
data set, the GR~orbit determination procedure forced systematically
small relative weights starting from January 2018. An NG~orbit can be
determined for this last data set, but the uncertainty of the
transverse component of the NG~acceleration is large. So the
conclusion is that between November 2013 and November 2014, at
heliocentric distances between 6.7\,au and 8.6\,au, C/2006~S3 still
exhibited NG effects in its orbital motion, while since November 2014,
they can be completely neglected. This, together with the previous
conclusions, indicates the rather erratic behavior of this comet about
two years after perihelion, which coincides perfectly with the
outburst event observed in this comet by \cite{sarneczky-et-al:2016}.

\begin{table}
\caption{\label{tab:2006s3_post_solution_series} Series of
  GR~solutions obtained for C/2006~S3 using three different sets of
  post-perihelion data arcs. Each includes the data from the years
  2018--2020. }

	\centering
	\setlength{\tabcolsep}{2.0pt} 
	\begin{tabular}{ccccccc}
		\hline 			\hline 
		\multicolumn{2}{c}{first observation} & \multicolumn{2}{c}{relative weights}  & 	\multicolumn{2}{c}{[O-C] trends} & RMS         \\
		date   & d$_h$    & Mauna Kea & DES  &  $\alpha$ & $\delta$                     & [arcsec]   \\
		\hline \\
		2015\,12\,10  & 10.59 & 1.38  & 4.45 & no  &  no   & 0\farcs 37  \\
		2014\,11\,27  & ~~8.57 & 1.34  & 4.38 & no  &  no   & 0\farcs 41 \\
		2013\,11\,21  & ~~6.72 & 0.22  & 0.32 & yes &  yes  & 0\farcs 48 \\
		\hline
	\end{tabular}
\end{table}

All of these GR and NG~orbits (based on different data arcs) lead to
$1/a_{\rm ori}$ values between slightly negative (for example, the
NG~orbit using the PRE data set gives $-0.49\pm 2.61$\,au$_{-6}$) and
about 20\,au$_{-6}$. NG~orbits in the MPC and the Nakano Notes give
values of 21.5\,au$_{-6}$ and 14\,au$_{-6}$, respectively, for
$1/a_{\rm ori}$.

\subsection{C/2008~S3 (Boattini)}

C/2008~S3 was discovered on 2008 September~29 when it was 9.91\,au
from the Sun.  Later, three series of pre-discovery measurements were
found: from 2007 December~18 (12.4\,au from the Sun), 2007 December~4
(11.0\,au), and 2006 December~27 (10.9\,au). Each series consists of
four measurements. Thus, the data arc covers 8.6\,yr in the
heliocentric distance range of 12.4\,au -- 8.01\,au (perihelion) --
11.7\,au.

For this comet, only GR~orbits could be determined. As a result, this
comet was included in the LBg sample. The CODE Catalog presents a
GR~orbit based on the complete data arc. This orbit fits the data with
an RMS of 0\farcs 42. The MPC and JPL also provide GR~orbits based on
the same data arc, fitted to positional data with RMS values of
0\farcs 60 and 0\farcs 73, respectively (as of July~2023).

In addition, we checked that all pre-discovery observations are also
well-fitted to orbits based on the data arc starting from the
discovery date. Here, we obtained GR~orbits based on the PRE and POST
data arcs with RMS values of 0\farcs 48 and 0\farcs 60, respectively.

GR orbits obtained using the complete and PRE data arcs result in
almost the same original $1/a$ of $21.0\pm 0.3$ and $21.3\pm 0.6$
\,au$_{-6}$. The GR~orbital solution presented in the MPC gives a
similar value of $21.0\pm 0.3$\,au$_{-6}$ for $1/a_{\rm ori}$.

\section{Original Values of Semimajor Axes}\label{sec:original-a}

Here, we give a detailed discussion of the values of the comets'
original semimajor axes. We focus on searching for correlations
between these values obtained in GR and NG~solutions based on the
three types of data arcs analyzed here: ~~(i) data before perihelion
(`PRE'), ~~(ii) data after perihelion (`POST'), and ~~(iii) all data
(`FULL').

\subsection{Sample LAn}\label{sec:original-a_LAn}

\begin{figure}
	\centering
\includegraphics[width=1.04\columnwidth]{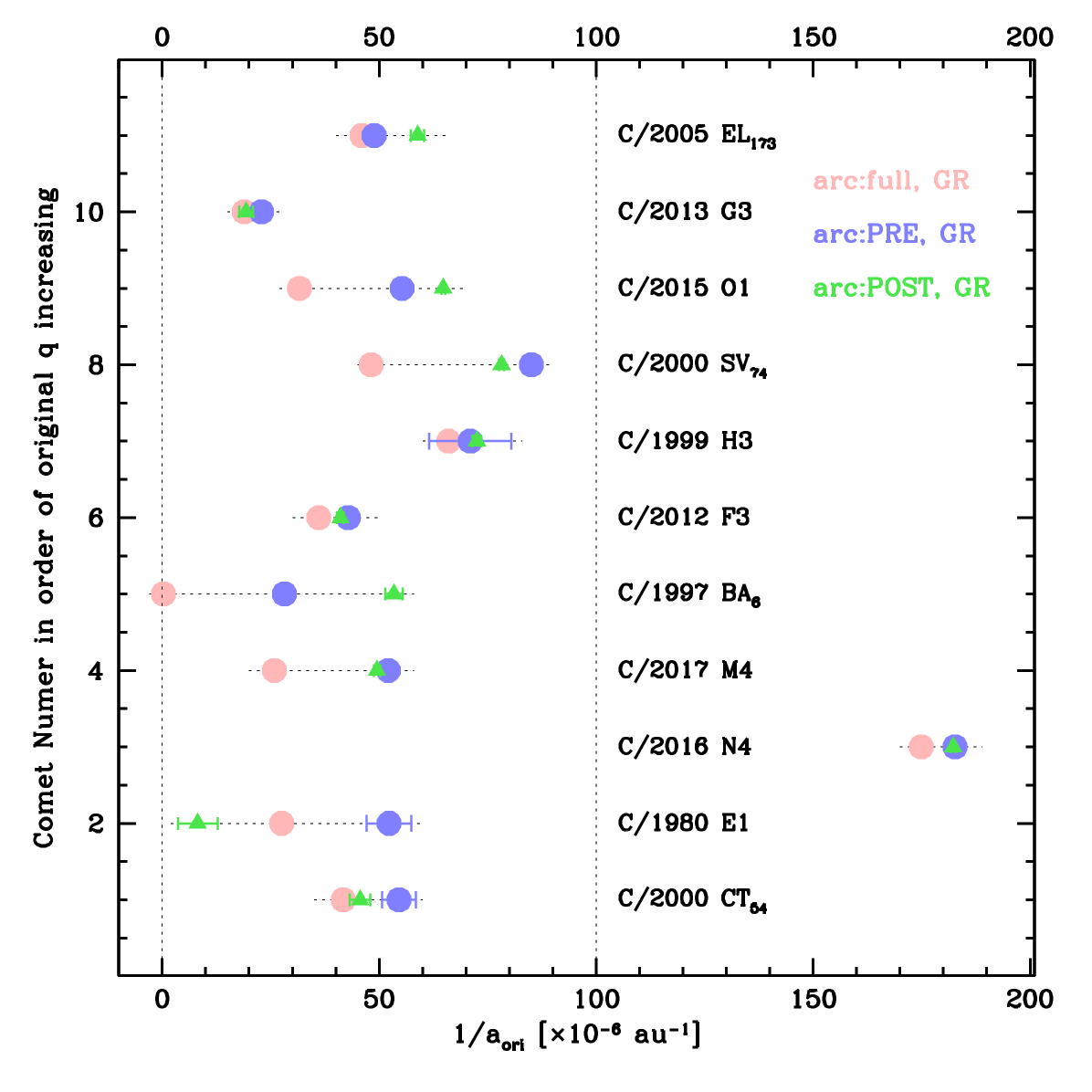}        
        \caption{Original $1/a$ values for the LAn sample using the
          whole data arc (light red dots, $1/a_{\rm ori,full,GR}$),
          pre-perihelion arcs (light blue, $1/a_{\rm ori,pre,GR}$),
          and post-perihelion arcs (light green, $1/a_{\rm
            ori,post,GR}$) based on purely gravitational orbital
          solutions. Comets with long data arcs, perihelion distances
          between 3.1\,au and 3.9\,au, and detectable NG effects using
          positional data are presented (Sample LAn). Comets are
          ordered on the vertical axis according to their perihelion
          distances; see Table~\ref{tab:comet_list}). 
            }
	\label{fig:LAn_gr}
\end{figure}

Values of $1/a_{\rm ori}$ for GR orbits derived using the three types
of data arcs are shown in Fig.~\ref{fig:LAn_gr} and are color-coded.
The red, blue, and light green marks represent orbits based on FULL,
PRE, and POST data arcs, respectively (also see
Table~\ref{tab:comet_list} for details of the data arcs). By
inspection of this figure, we note the following tendencies for GR
orbits.

\begin{enumerate}
\item For all comets, $1/a_{\rm ori,full,GR} < 1/a_{\rm
  ori,pre,GR}$. Thus $a_{\rm ori,full,GR} > a_{\rm ori,pre,GR}$, so
  the original semimajor axis inferred from the FULL data arc is
  larger than that derived from the PRE data arc. For the two comets
  with the largest perihelion distances, C/2005~EL$_{173}$ and
  C/2013~G3, as well as for C/1999~H3, the PRE and FULL values are
  close to each other (they differ by less than 10\,au$_{-6}$). For
  the other comets in this sample, the $1/a_{\rm ori}$ differences are
  as large as 30-40\,au$_{-6}$.
\item In ten of the eleven cases, we also have $1/a_{\rm ori,full,GR}
  < 1/a_{\rm ori,post,GR}$. For comets C/2013~G3 and C/2000~CT$_{54}$,
  the POST and FULL values are similar. Only C/1980~E1 has $1/a_{\rm
    ori,full,GR} > 1/a_{\rm ori,post,GR}$.
\item Points 1 and 2 show that in ten of the eleven cases, the
  original semimajor axis based on the FULL data arc is larger than
  the semimajor axis obtained using either the PRE or POST data.
  These systematic trends in GR~fits are so large that they cannot be
  explained by the uncertainties of orbits determined from different
  data sets. They must be related to the NG~accelerations acting in
  the motion of these comets.
\item For four of the eleven cases (C/2017~M4, C/2016~N4, C/2012~F3,
  C/1999~H3), $1/a_{\rm ori,pre,GR} \simeq 1/a_{\rm
    ori,post,GR}$. However, both values differ substantially from
  $1/a_{\rm ori,full,GR}$.
\item For seven of the eleven comets, the original semimajor axis
  inferred from the PRE data arc is smaller than the values from both
  the POST and FULL data arcs: $1/a_{\rm ori,pre,GR} > 1/a_{\rm
    ori,full,GR}$ and $1/a_{\rm ori,pre,GR} > 1/a_{\rm ori,post,GR}$.
\end{enumerate}

\begin{figure}
	\centering
\includegraphics[width=1.04\columnwidth]{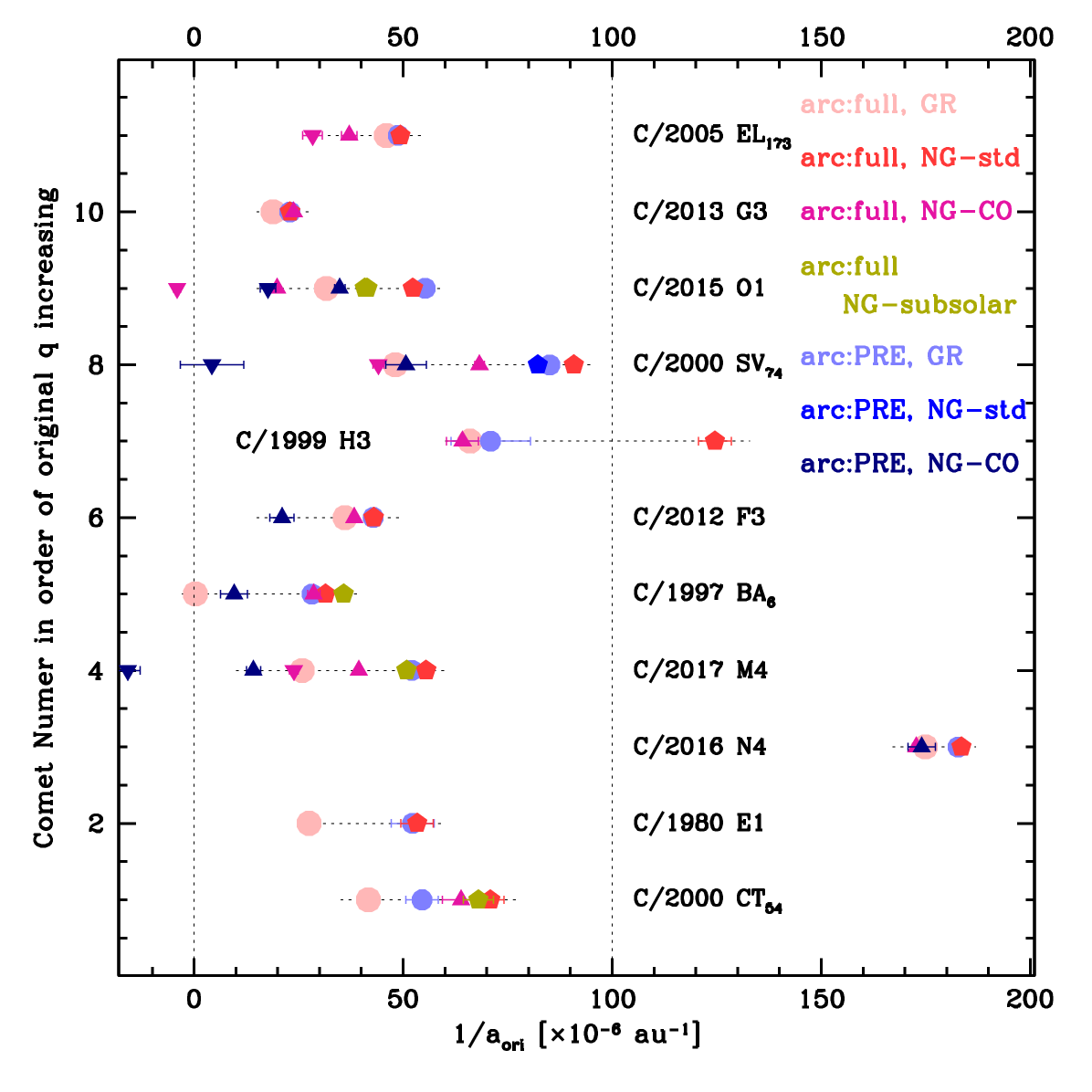}        
	\caption{Original $1/a$ values for the same comets as in
          Fig.~\ref{fig:LAn_gr} using FULL and PRE data arcs and both
          GR and a variety of NG~solutions, where NG~solutions
          obtained using the whole data arcs are shown by red
          pentagons (standard $g(r)$, $1/a_{\rm ori,full,NG-std}$),
          and magenta triangles ($g(r)$-like function based on CO
          sublimation, $1/a_{\rm ori,full,NG-CO}$), where
          upward-pointing and inverted triangles show orbits using
          $r_0=10$\,au and $r_0=50$\,au, respectively. Inverted
          triangles are given for four comets only; for more
          discussion, see Secs.~\ref{sec:2000sv_2015o1} and
          \ref{sec:original-a_LAn}.  }
	\label{fig:LAn_gr_ngfull}
\end{figure}

If we include NG solutions for this analysis, then the relations
between $1/a_{\rm ori}$ values will depend on the assumed form of the
$g(r)$-like function.

Fig.~\ref{fig:LAn_gr_ngfull} shows the differences between $1/a_{\rm
  ori}$ resulting from the NG~orbit based on the standard form of the
$g(r)$ function (red and blue pentagons) and GR solutions (light red
and light blue dots). First, the differences in original $1/a$ between
GR and NG-std solutions based on the FULL data arc can reach
50\,au$_{-6}$.  It is also noticeable that $1/a_{\rm ori,full, GR} <
1/a_{\rm ori,full,NGstd}$.  This systemic trend of $1/a_{\rm ori}$
being larger for NG orbits is well-known; see \cite{squires:1961,
  marsden-sek-ye:1973, mar-sek-eve:1978, krolikowska:2001,
  yeomans_cho_sit_szut_krol:2004, rickman-2014,
  krolikowska:2020_NG}. \citet{mar-sek-eve:1978} found that for a
sample of near-parabolic comets, most with perihelion distances $q <
2$~au, NG forces had a large effect on the inferred original semimajor
axes. For a comet with $q < 0.5$~au, for example, a purely
gravitational fit would typically yield a hyperbolic orbit (their
Eq.~2).  \cite{krolikowska:2001} studied 33 comets whose original GR
orbits were slightly hyperbolic, i.e., $1/a_{\rm ori, GR} < 0$. For
sixteen of the comets, she was able to fit NG orbits. In every case,
$1/a_{\rm ori}$ increased when NG effects were accounted for. Fourteen
of the orbits changed from hyperbolic to elliptical, while the other
two remained hyperbolic at the $\approx 2 \sigma$ level.

The further analysis in this paper suggests that this correlation is
related to the standard and HJ17 forms of the $g(r)$-like
functions. For the NG-CO form of the $g(r)$-like function studied
here, this systematic effect seems to be weaker (see below).

However, we find other correlations. In all cases, the GR~solution
based on the FULL data arc gives values of $1/a_{\rm ori}$ (light red
dots in the figure) smaller than either the NG solution obtained using
the FULL data arc (red pentagons) or the GR solution based on the PRE
data arc (light blue dots).  In other words, the GR orbit obtained
with the FULL data arc results in the largest original semimajor axis
of these three solutions.

In addition, in ten of eleven cases, the original semimajor axes
satisfy the relationship $1/a_{\rm ori,full, GR} < 1/a_{\rm ori,PRE,
  GR} < 1/a_{\rm ori,full,NGstd}$.  The exception is C/2015~O1, for
which $1/a_{\rm ori,full, GR} < 1/a_{\rm ori,full,NGstd} < 1/a_{\rm
  ori,PRE, GR}$.  Moreover, for eight of the comets, the red pentagons
and light blue dots coincide (C/1980~E1, C/2016~N4, C/2012~F3,
C/2013~G3, C/2005~EL$_{173}$) or are very close to each other
(C/2017~M4, C/1997~BA$_6$, C/2015~O1). For these comets, the GR~orbit
based on the PRE data arc gives a very similar original semimajor axis
(differences in $1/a_{\rm ori}$ are less than 5\,au$_{-6}$) as in the
case of the NG~solution based on the FULL data arc and standard $g(r)$
form.  Thus, an analysis of these three types of solutions shows that
the value of $1/a_{\rm ori,GR}$ based on the PRE~solution is in all
cases closer to the NG-std solution based on the FULL data arc than
the value based on the GR orbit derived from the FULL data arc. In
this way, we confirm the intuitive conclusion that the GR~solution
based on the PRE data arc is better suited for studying the origin of
comets with perihelion distances $q<4$\,au, as well as when it is not
possible to reliably determine NG~orbits from the FULL data arc (when
assuming the standard form of the $g(r)$ function).

Assuming the $g(r)$-like function of HJ17 for sublimation of water ice
from the subsolar point, the values of $1/a_{\rm ori}$ for most comets
analyzed here are quite similar to those for the NG-std
case. Fig.~\ref{fig:LAn_gr_ngfull} shows values of $1/a_{\rm ori}$ for
the $g(r)$-like form describing sublimation from the subsolar point
for four comets (gold points), where for three of them (C/2017~M4,
C/1997~BA$_6$, and C/2015~O1) this type of NG~solution gives the
smallest RMS between the three forms of NG~solutions based on the
complete data sets (see Table~\ref{tab:NG_solutions}).  They, however,
represent a typical situation for the relative location of $1/a_{\rm
  ori,full, NGstd}$ and $1/a_{\rm ori,full,NGsubsolar}$. {Among the
  remaining NG~orbits based on HJ17, only in the case of comet
  C/1999~H3 is the value of $1/a_{\rm ori, full, NGsubsolar}$
  significantly smaller than $1/a_{\rm ori, full, NGstd}$ and is equal
  to $89.53\pm 3.78$\,au$_{-6}$ (not given in the figure because the
  RMS is worse than for the NG~orbit based on the standard
  $g(r)$~formula)}.

\vspace{0.4cm}

For the $g(r)$-like function describing CO ice sublimation, the
situation is different from water ice sublimation.
Fig.~\ref{fig:LAn_gr_ngfull} also shows $1/a_{\rm ori}$ for the NG-CO
solution based on the FULL data arc (magenta triangles) and the PRE
data arc (dark blue triangles) for comets from the sample LAn. For at
least half of the cases, this form of the $g(r)$-like function gives a
slightly worse orbital fit to the data arc
(Sec.~\ref{sec:2000sv_2015o1}).  The positions of the red, gold, and
magenta marks show that the value of $1/a_{\rm ori}$ depends on the
assumed form of the $g(r)$-like function. In addition, there is not
such a clear systematic effect for NG-CO orbits as there was for the
NG orbits assuming water sublimation. The relationship $1/a_{\rm
  ori,full,GR} > 1/a_{\rm ori,full,NG_{CO}}$ applies to four of the
eleven comets: C/2016~N4, C/1999~H3, C/2015~O1, and C/2005~EL$_{173}$
when we used the NG-CO form with $r_0=10$\,au. However, for all four
comets for which we tested $r_0=50$\,au (inverted triangles in the
figure), the values of $1/a_{\rm ori}$ are notably smaller than those
using the NG-CO form with $r_0=10$\,au. The values of $1/a_{\rm ori}$
obtained using $r_0=50$\,au are smaller than $1/a_{\rm ori,full,GR}$
for the next two comets (C/2017~M4 and C/2000~SV$_{74}$). Note that
for these two comets, assuming $r_0 = 50$~au gives slightly hyperbolic
original orbits.

\subsection{Sample LBn}\label{sub:LBn}

For comets with $q>4.5$\,au, only the $g(r)$-like form that describes
CO-driven sublimation is analyzed.

\begin{figure}
	\centering
	\includegraphics[width=1.04\columnwidth]{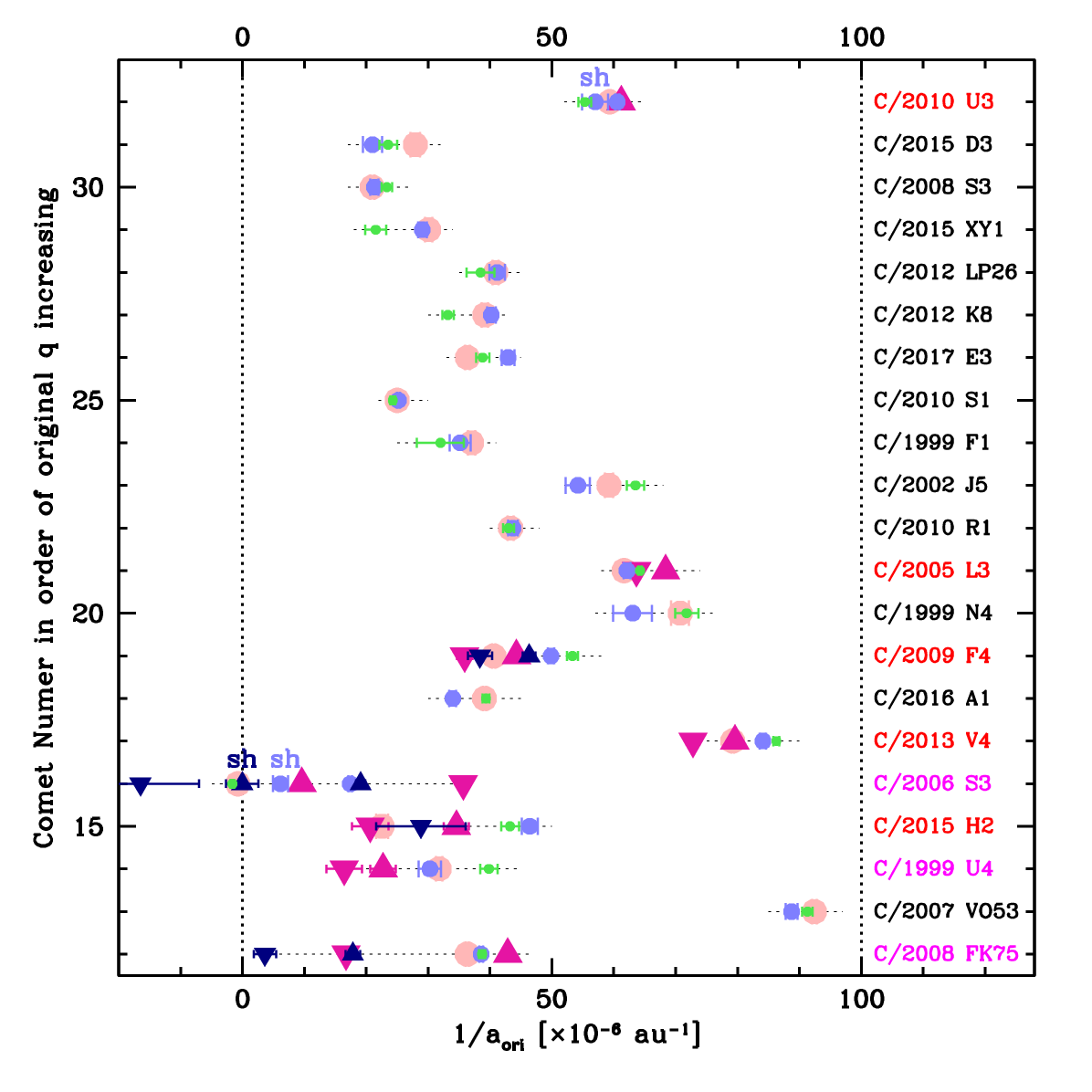}
\caption{Original $1/a$ for comets from samples LBn (marked by red and
  magenta labels on the right side of the figure) and LBg (black
  labels). Comets are ordered on the vertical axis according to their
  perihelion distances. Three types of marks representing GR orbits
  are color-coded as in Fig.~\ref{fig:LAn_gr_ngfull}. NG~solutions
  (triangles) are obtained using the $g(r)$-like function representing
  CO sublimation. Upward- and downward-pointing magenta triangles show
  orbits based on the whole data arc using $r_0=10$\,au and
  $r_0=50$\,au, respectively; small blue triangles indicate orbits
  using the PRE data arc; `sh' means solutions based on the distant
  part of the pre-perihelion data arc. On the right side of the
  figure, red names show comets with NG~orbits determined here;
  magenta names indicate comets with NG~orbits obtained earlier; black
  names show comets with indeterminable NG effects using positional
  data (sample LBg).}
	\label{fig:LBg_LBn}
	
\end{figure}

Fig.~\ref{fig:LBg_LBn} includes $1/a_{\rm ori}$ obtained for LPCs in
the samples LBn and LBg. The red and magenta labels given on the right
side of this plot indicate comets from the LBn sample. Those marked in
red are comets with NG~orbits obtained in the present study, while
those in magenta are the remaining comets of the Sample~LBn.

For seven comets in Sample~LBn (all except C/2010~U3, which gave an
unphysical solution with $A_1<0$), we calculated $1/a_{\rm ori}$ using
a $g(r)$-like function for CO, but with $r_0 =$~50~au. Those points
are shown as inverted triangles. Using this value of $r_0$ does not
improve the RMS or [O-C] of the fits; in two cases, the RMS is worse,
see Table~\ref{tab:NG_solutions}.

Seven of the eight comets in this sample fulfill the relationship
$1/a_{\rm ori,full,GR} < 1/a_{\rm ori,pre,GR}$; the reverse applies
only for C/1999~U4. In the case of C/2010~U3, the figure shows two
GR~solutions based on the PRE data arcs (light blue points). The light
blue point with the `sh' label indicates the solution based on the
data arc limited to distances greater than 14\,au from the Sun.  In
this case, $1/a_{\rm ori}$ is also smaller than the value resulting
from a GR~orbit based on a FULL data arc.

For six of eight LPCs, we have $1/a_{\rm ori,full,GR} < 1/a_{\rm
  ori,post,GR}$, while the opposite is true for
C/2006~S3 and C/2010~U3. However, for these two comets, both values of $1/a_{\rm
  ori}$ are very close to each other.

When considering all three data arcs, for five of eight LPCs, the
GR~orbit based on the FULL data arc gives a smaller value of $1/a_{\rm
  ori}$ than the other two GR~solutions (based on PRE and POST data
arcs). For comets C/1999~U4, C/2006~S3, and C/2010~U3, this
relationship does not hold.

In other words, although we observe some relationships between the
values of $1/a_{\rm ori}$ for different types of GR~orbits, the trends
are weaker than they are for the LAn sample.  This may be a
consequence of the generally weaker NG effects for comets in the LBn
sample than in the LAn sample.  In this sample, all values of
$1/a_{\rm ori}$ for the three GR solutions (PRE, FULL, and POST data
arcs) do not differ by more than 15\,au$_{-6}$ for six comets. The
difference between $1/a_{\rm ori}$ GR~orbits is smaller than
25\,au$_{-6}$ for the two remaining comets -- C/2015~H2, which shows
the strongest evidence of NG~acceleration in the motion of the sample
LBn, and C/2006~S3. However, when an orbit based on a distant part of
the PRE data arc was considered (light blue point labeled as `sh' in
Fig~\ref{fig:LBg_LBn}), instead of an orbit based on the complete PRE
data arc, this range of $1/a_{\rm ori}$ decreased to about
10\,au$_{-6}$ for C/2006~S3 (GR and NG~solutions based on full data
and PRE+DIST data).

For six of seven comets with NG-CO~orbits obtained using full data
arcs and $r_0=50$\,au (inverted magenta triangles in the figure), the
values of $1/a_{\rm ori}$ are significantly smaller than those using
the NG-CO form with $r_0=10$\,au (only for C/2006~S3 is the reverse),
and in five cases the values of $1/a_{\rm ori}$ obtained with
$r_0=50$\,au are smaller than $1/a_{\rm ori,full,GR}$. Additionally,
we even find a negative value of $1/a_{\rm ori}$ for C/2006~S3 when
the PRE data set is used.

\subsection{Sample LBg} \label{sub:LBg}

The NG~effects in the motion of thirteen of thirty-two LPCs analyzed
here were indeterminable or very poorly determined using their
positional data sets.  There are some signs of NG effects for
C/1999~N4 and C/2012~LP$_{26}$, but they are highly uncertain.

The black labels in Fig~\ref{fig:LBg_LBn} indicate the LBg sample
comets.  For six of the thirteen LPCs (C/2010~R1, C/1999~F1,
C/2010~S1, C/2012~K8, C/2012~LP$_{26}$, C/2015 XY1, and C/2008~S3),
the values of $1/a_{\rm ori,full,GR}$ (light red dots) and $1/a_{\rm
  ori,pre,GR}$ (light blue dots) are very close to each other (values
differ by no more than 2\,au$_{-6}$). In all of these cases, $1/a_{\rm
  ori,post,GR}$, shown with a light green point, also has a very
similar value but typically is more different than the other two. The
LBn sample contains three cases like this (C/2008~FK$_{75}$,
C/2005~L3, and C/2010~U3).

Unlike the previous two samples, as many as seven comets in sample LBg
exhibit the opposite relationship $1/a_{\rm ori,pre,GR} < 1/a_{\rm
  ori,full,GR}$. The $1/a_{\rm ori,pre,GR} > 1/a_{\rm ori,full,GR}$
trend observed in the LAn and LBn samples thus does not apply here.

In addition, the LBg sample contains as many as nine comets with
$1/a_{\rm ori,post,GR} < 1/a_{\rm ori,full,GR}$, although the
difference between both values is very small for three of these
comets.  We tentatively conclude that determining the orbit solely
from the POST data arc will typically result in original orbits with a
greater semimajor axis than the comet has in reality.

In general, however, in the LBg sample, all three values of $1/a_{\rm
  ori}$ for GR~solutions (PRE, FULL, and POST data arcs) do not differ
by more than 15\,au$_{-6}$.

\section{Final Remarks and Conclusions}\label{sec:conclusions}

\begin{figure}
	\centering
	\includegraphics[width=1.04\columnwidth]{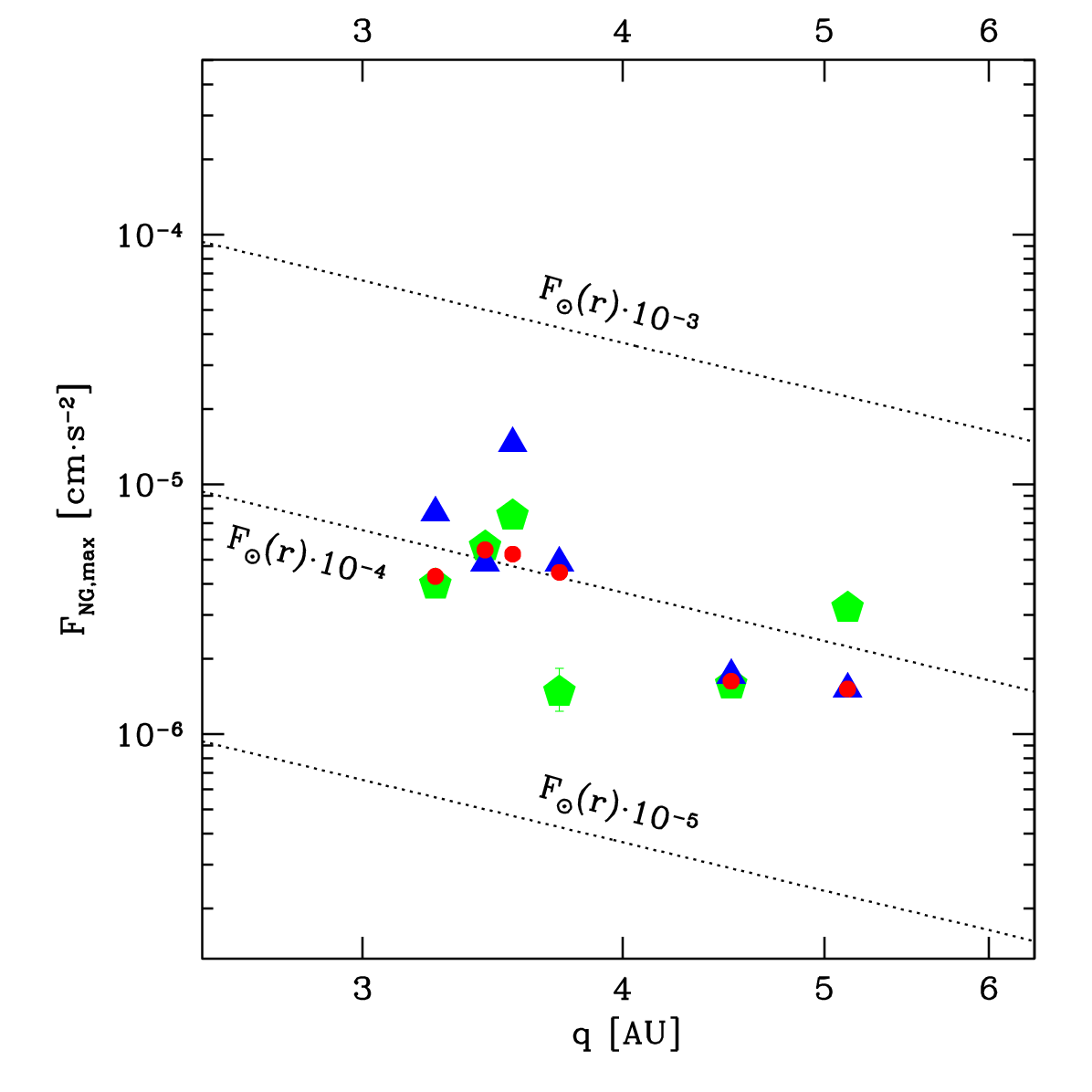}
	\caption{Strength of NG acceleration acting on cometary nuclei
          at perihelion for six comets with determined NG~parameters
          using PRE and POST perihelion data arcs, separately; both
          axes are on logarithmic scales. The value of $F_{\rm
            NG,max}$ is calculated as: $\sqrt{ {A_1}^2 + {A_2}^2 +
            {A_3}^2} \cdot g(q)$. The blue triangles represent
          NG~orbits using PRE data sets; green pentagons, POST data
          sets; and red dots, NG~orbits based on the entire data
          sets. All use the $g(r)$-like function for CO with $r_0 =
          10$~au. From the left to right, the comets shown are
          C/2017~M4, C/1997~BA$_6$, C/2000~SV$_{74}$, C/2015~O1,
          C/2008~FK$_{75}$, and C/2006~S3.  The three black dotted
          lines show $10^{-3}$, $10^{-4}$, and $10^{-5}$ of $F_{\sun}
          (r)$, where $F_{\sun} (r) = 0.59\cdot r^{-2}$ is the solar
          gravitational acceleration in units of cm s$^{-2}$ if $r$ is
          given in au.  }
	\label{fig:FNmax}
\end{figure}

This study shows that the motion of comets can often be measurably
affected by NG forces at heliocentric distances well beyond
5\,au. However, it is difficult to accurately model NG~accelerations
far from the Sun (Table~\ref{tab:NG_parameters} and
Fig.~\ref{fig:LBg_LBn}). NG~effects are also a source of additional
uncertainties resulting from the adopted form of the $g(r)$-like
function (Fig.~\ref{fig:LAn_gr_ngfull}). Therefore, the orbital
changes caused by the NG acceleration far from the Sun constitute a
serious obstacle in dynamical studies of the origin of comets on
near-parabolic orbits, i.e., comets with eccentricities very close to
1 (Fig.~\ref{fig:LBg_LBn}, also see
\citet{jewitt:2021_distant_activity_2017k2}). In such cases,
determining whether a comet is dynamically new or old requires an
individual approach for each object
(Sec.~\ref{sec:2000sv_2015o1}--\ref{sec:other_examples}).

In general, when the PRE arc includes data at large heliocentric
distances, the original $1/a$ uncertainty associated with the NG
effects is minimized. Unfortunately, such observations are not
available for all comets.  In practice, it is necessary for each comet
to balance the benefits of limiting the action of NG~acceleration (the
farther from the Sun, the better) and the quality of the obtained
orbit (the longer the data arc, the better). This paper shows how this
works for comets found at about 7\,au from the Sun. Fortunately, the
number of LPCs discovered at even larger distances, such as comet
C/2010~U3 (large $q \approx 8.4$\,au, see
Table~\ref{tab:Oortspikecomets10au}) analyzed here, or C/2017~K2
(small $q \approx 1.8$\,au) discussed in
\citet{jewitt:2021_distant_activity_2017k2} and
\citet{dyb-kroli:2022}, will increase in the future.

Another interesting issue is the usefulness of data recorded well
before the discovery of the comet to test the existence of traces of
NG acceleration in the motion of the comet far from the Sun. The two
prominent cases discussed in this study with such data are C/2010~U3
and C/2006~S3. Pre-discovery data proved to be important in orbit
determination for C/2010~U3. However, for C/2006~S3, the earliest
pre-discovery series of data could not be integrated into the models
of motion applied here. The failure to fit C/2006~S3 may be due to
some systematic error in the measured positions. However, it cannot be
ruled out that the NG models used here are inadequate for this comet
because, for example, a fragment broke away from the comet's nucleus
long before perihelion or the comet underwent an outburst. Such an
explanation, as one of many possibilities, seems potentially much more
interesting in the context of constructing the NG~model of motion for
this comet.

\vspace{0.2cm}

Other general conclusions based on the analysis carried out in this
research for LPCs with large perihelion distances are as follows.

\begin{itemize}
\item For 19 of 32 LPCs discovered far from the Sun, there are notable
  trends in fitting the GR orbit within the data arc, which most
  likely indicate the action of NG forces. We determined NG orbits of
  six comets for the first time. We discuss the NG orbit fitting to
  the data in comparison to the purely GR solutions
  (Sec.~\ref{sec:five_comets}).  The conclusion is that more refined
  NG models of motion should be constructed individually for each
  comet using photometry or spectroscopic observations if possible. We
  complete the discussion of new NG~solutions in comets with large
  perihelia with several other examples, already known, such as
  C/2006~S3 (Sec.~\ref{sec:other_examples}). However, such comparisons
  have never been shown in the literature before.
  
\item To calculate the past evolution of near-parabolic comets, even
  those with large perihelion distances are better studied using their
  orbits determined based on pre-perihelion data (the PRE data arc).
  For LPCs discovered far from the Sun, the data arc can even be
  limited to part of the pre-perihelion leg of the orbit. However, the
  orbit should remain of the highest quality. This is the recommended
  direction for examining the origin of LPCs, already suggested in our
  previous research \citep{kroli-dyb:2012, krolikowska:2020_NG}.
  Nowadays, more comets are being discovered far from the Sun, so
  there will soon be more cases of this kind.

\item For three types of GR orbits (based on PRE, FULL, and POST data
  arcs), $1/a_{\rm ori}$ can differ by up to 50\,au$_{-6}$ for LPCs
  with 3\,au~$<q<$~4\,au (sample LAn). NG~orbits based on different
  forms of the $g(r)$-like function result in similar differences in
  $1/a_{\rm ori}$ for comets in this range of $q$ (see
  Fig.~\ref{fig:LAn_gr_ngfull}).  For LPCs with perihelia $> 4.5$\,au,
  these differences between GR~orbits are always less than
  25\,au$_{-6}$ (sample LBn), and in most cases less than
  15\,au$_{-6}$ (samples LBg and LBn, see
  Fig.~\ref{fig:LBg_LBn}). Differences at this level are of
  fundamental importance in the study of the dynamical status of these
  objects and their origin. In this regard, particular attention
  should be paid to comets discovered after perihelion
  passage. Especially for such LPCs, finding pre-discovery
  observations is invaluable.
  
\item In cases in which the NG~effects can be determined, the value of
  $1/a_{\rm ori}$ based on NG~orbits using the standard form of the
  $g(r)$ function and the FULL data arc is usually more compatible
  with the $1/a_{\rm ori}$ from the GR~orbits determined from the PRE
  data arc than with the $1/a_{\rm ori}$ for the other types of
  GR~orbits (POST and FULL data arcs, sample LAn). Thus, when it is
  impossible to determine the NG orbit and the PRE data arc results in
  a GR orbit of high quality, it is worth using the PRE orbit to study
  the past evolution of LPCs, instead of using the GR orbit fitted to
  the FULL data arc.
  
\item These studies indicate that the systematic effect of smaller
  original semimajor axes for the NG~orbits than the GR~orbits
  \citep{marsden-sek-ye:1973, mar-sek-eve:1978, krolikowska:2001,
    yeomans_cho_sit_szut_krol:2004, rickman-2014, krolikowska:2020_NG}
  seems to be related to the specific form of the
  \cite{marsden-sek-ye:1973} $g(r)$ function describing water-ice
  sublimation, in particular to the heliocentric position of the
  `knee' of this function (see Fig.~\ref{fig:grlike}). When using the
  $g(r)$-like form with a greater value of $r_0$ (for example
  dedicated to CO sublimation, that is to say, comets with large
  perihelion distances as analyzed here), this systematic effect may
  not be so conspicuous (Sec.~\ref{sec:original-a}). We intend to
  investigate this issue in the future.

The $g(r)$ form of \cite{marsden-sek-ye:1973} is a fit to a very
simple thermal model for a spherical cometary nucleus which assumes a
single average value for the solar elevation angle seen by the
nucleus, and neglects heat transport into the comet (Sec.~1; see,
e.g., \cite{Huebner2006}). Rosetta's encounter with comet 67P has
shown the limitations of the \cite{marsden-sek-ye:1973} model. More
elaborate NG force models exist \citep{sitarski:1990, Sekanina1993a,
  Sekanina1993b, Maquet2012}, but such models have more free
parameters and cannot yet be applied to a sufficiently large sample of
near-parabolic comets for an analysis of a general
nature\footnote{Some such models only apply to comets seen on more
than one apparition, and so cannot be applied to near-parabolic comets
at all.}.  However, it seems to us that using subsets of the data arc,
a variety of expressions for $g(r)$, and, when available, measured
sublimation rates \citep{krolikowska:2004} is a promising direction
for future study of Oort Cloud comets with the longest observation
arcs and NG~effects clearly visible in the positional data.
 
\item Orbit fits provided by JPL for comets assumed to have CO-driven
  activity set the knee in $g(r)$ at $r_0 = 5$~au. Both experimental
  data \citep{Fray2009, Luna2014} and activity of comets beyond 20~au
  (e.g., \citet{yang:2021_CO_2017k2}) imply that the real value of
  $r_0$ must be larger. Here we have assumed $r_0 = 10$~au for
  CO-driven activity in most cases, but have also tried $r_0 = 15$,
  30, and 50~au for some comets. In terms of RMS and [O-C], these fits
  are generally of similar quality to those with $r_0 = 10$~au, but
  the inferred values of $a_{\rm ori}$ are surprisingly sensitive to
  the assumed $r_0$. We will investigate this issue in future work.
 
\item Our analysis of NG~effects based on positional data suggests
  that these effects can be substantially different before and after
  perihelion. Unfortunately, we were able to determine reliable
  NG~orbits based on PRE and POST data for only six comets. We found
  that for two comets, their NG parameters were larger after
  perihelion (four times bigger for C/2006~S6 and 30\% larger for
  C/1997~BA$_6$), and for three others, larger before perihelion
  (almost two times bigger for C/2017~M4, two times for C/2015~O1, and
  three times for C/2000~SV$_{74}$), where the asymmetry was measured
  by comparing the value of $A = \sqrt{ {A_1}^2 + {A_2}^2 + {A_3}^2}$
  for the NG~solution using CO sublimation. For the sixth comet,
  C/2008~FK$_{75}$, the level of NG~parameters before and after
  perihelion is comparable (the asymmetry is less than 10\%); see
  Fig.~\ref{fig:FNmax}. However, the relative difference in
  NG~activity depends on the assumed form of the $g(r)$-like function
  (Sec.~\ref{subsec:2017m4}).
\end{itemize}

Finally, it is worth highlighting that pre-perihelion observations
covering a wide range of large heliocentric distances allow us to
accurately determine the original orbits of comets without the major
uncertainties related to the existence of NG acceleration in the data
arc considered.  In the future, we will have to deal with such
favorable cases more often as we discover more and more comets beyond
Saturn's orbit and even further from the Sun.

\begin{acknowledgements}
	This research has made use of positional data of comets
        provided by the International Astronomical Union's Minor
        Planet Center. We thank M\'ario De Pr\'a, Piotr Dybczy\'nski,
        Arika Higuchi, David Nesvorn\'y, Scott Tremaine, and Maria
        Womack for discussions and Marc Fouchard for a helpful review.
\end{acknowledgements}

\bibliography{Oort_long_data_arcs_2023august7}

\end{document}